\documentclass{aa} 
\usepackage{graphicx}
\usepackage{natbib}
\usepackage[utf8]{inputenc}
\usepackage{graphics}
\usepackage[subnum]{cases}
\usepackage{threeparttable}
\usepackage[titletoc,title]{appendix}
\usepackage[dotinlabels]{titletoc}
\usepackage[colorlinks=true,linkcolor=black,citecolor=blue]{hyperref}
\usepackage{makecell}
\usepackage{lscape}
\usepackage[figuresright]{rotating}
\usepackage{caption}
\usepackage{multirow,booktabs}
\usepackage{ulem}
\usepackage{txfonts}
\usepackage{colortbl}                          
\usepackage{ulem}
\usepackage{textcomp}
\usepackage{fontenc}
\usepackage{xspace}
\begin{document}

\title{Electron density distribution and solar plasma correction  \\
of radio signals using MGS, MEX and VEX spacecraft navigation data and its application to planetary ephemerides}

 \author{A. K. Verma\inst{1,2} \and A. Fienga\inst{1,3} \and J. Laskar\inst{3} \and K. Issautier\inst{4} \and H. Manche\inst{3} \and M. Gastineau\inst{3}
}

 \institute{ Observatoire de Besan\c con, CNRS UMR6213,
 41bis Av. de l'Observatoire, 25000 Besan\c con, France
  \and
 CNES, Toulouse, France
 \and
 Astronomie et Syst\`emes Dynamiques,
 IMCCE-CNRS UMR8028,
 77 Av. Denfert-Rochereau, 75014 Paris, France
 \and
 LESIA, Observatoire de Paris,CNRS, UPMC, Université Paris Diderot, 5 Place Jules Janssen, 92195 Meudon, France
 }

 \offprints{A. Verma, ashok@obs-besancon.fr}

 \date{\today}

 \titlerunning{Electron density distribution from MGS, MEX and VEX}
 \authorrunning{Verma et al}

  \abstract{The Mars Global Surveyor (MGS), Mars Express (MEX), and Venus Express (VEX) experienced several  superior solar conjunctions. These conjunctions cause severe degradations of radio signals when the line of sight between the Earth and the spacecraft passes near to the solar corona region. 
  The primary objective of this work is to deduce a solar corona model from the spacecraft navigation data acquired at the time of solar conjunctions and to estimate its average electron density. The corrected or improved data are then used to fit the dynamical modeling of the planet motions, called planetary ephemerides.
 We analyzed the radio science raw data of the MGS spacecraft using the orbit determination software GINS. The range bias, obtained from GINS and provided by ESA for MEX and VEX, are then used to derive the electron density profile. These profiles are obtained for different intervals of solar distances: from 12R$_\odot$ to 215R$_\odot$ for  MGS, 6R$_\odot$ to 152R$_\odot$ for MEX, and form 12R$_\odot$ to 154R$_\odot$ for VEX. They are acquired for each spacecraft individually, for ingress and egress phases separately and both phases together, for different types of solar winds (fast, slow), and for solar activity phases (minimum, maximum).
 We compared our results with the previous estimations that were based on {\itshape in situ} measurements, and on solar type III radio and radio science studies made at different phases of solar activity and at different solar wind states. Our results are consistent with estimations obtained by these different methods. 
Moreover, fitting the planetary ephemerides including complementary data that were corrected for the solar corona perturbations, noticeably improves the extrapolation capability of the planetary ephemerides and the estimation of the asteroids masses.
 
  }
 
  \maketitle


\section{Introduction}
The solar corona and the solar wind contain primarily ionized hydrogen ions, helium ions, and electrons. These ionized particles are ejected radially from the Sun. The solar wind parameters, the velocity, and the electron density are changing with time, radial distances (outward from Sun), and the solar cycles \citep{Schwennvol1,Schwennvol2}. The strongly turbulent and ionized gases within the corona severely degrade the radio wave signals that propagate between spacecraft and Earth tracking stations. These degradations cause a delay and a greater dispersion of the radio signals. The group and phase delays induced by the Sun activity are directly proportional to the total electron contents along the {LOS} and inversely proportional with the square of carrier radio wave frequency.

\begin{table}
\caption{Previous models based on {\itshape in situ} and radio science measurements (see text for detailed descriptions).}
\centering
\small
\begin{tabular}{ l l l }
  \hline  
                        &                           &                                            \\
    Spacecraft       &    Data type   & Author  \\
 \hline

    Mariner 6 and 7 &radio science&   \cite{Muhleman77}\\
    Voyager 2 & radio science&   \cite{AndersonV2} \\

    Ulysses & radio science &   \cite{Bird94} \\

        Helios land 2 & {\itshape in situ} &   \cite{Bougeret}\\

     Ulysses & {\itshape in situ} &   \cite{Issautier98}\\
     Skylab& Coronagraph & \cite{Guhathakurta96} \\
     Wind &solar radio  &   \cite{Leblanc}\\
    & burst III & \\
    \hline
    
\end{tabular}
\label{literature}
\end{table}

By analyzing spacecraft radio waves facing a solar conjunction (when the Sun directly intercepts the radio signals between the spacecraft and the Earth), it is possible to study the electron content and to better understand the Sun. An accurate determination of the electron density profile in the solar corona and in the solar wind is indeed essential for understanding the energy transport in collision-less plasma, which is still an open question \citep{Cranmer02}. Nowadays, mainly radio scintillation and white-light coronagraph measurements can provide an estimation of the electron density profile in the corona \citep{Guhathakurta94,Bird94,Guhathakurta96,Guhathakurta99,Woo99}. However, the solar wind acceleration and the corona heating take place between 1 to 10 R$_\odot$ where {\itshape in situ} observations are not possible. Several density profiles of solar corona model based on different types of data are described in the literature (Table \ref{literature}). The two viable methods that are generally used to derive these profiles are \citep{Muhleman81} (1) direct {\itshape in situ} measurements of the electron density, speed, and energies of the electron and photons, and (2) an analysis of single- and dual-frequency time delay data acquired from interplanetary spacecraft. 

We performed such estimations using {MGS}, {MEX}, and {VEX} navigation data obtained from 2002 to 2011. These spacecraft  experienced several  superior solar conjunctions. This happened for MGS in 2002 (solar activity maximum), for MEX in 2006, 2008, 2010, and 2011 (solar activity minimum), and for VEX in 2006, 2008 (solar activity minimum). The influences of these conjunctions on a spacecraft orbit are severely noticed in the post-fit range and can be seen in the Doppler residuals obtained from the orbit determination software (see Figure \ref{DOP}). 

In section 2, we use Doppler- and range-tracking observations to compute the MGS orbits. From these orbit determinations, we obtained range systematic effects induced by the planetary ephemeris uncertainties, which is also called range bias. For the MEX and VEX spacecraft, these range biases are provided by the {ESA} \citep{Fienga2009}. These range biases are used in the planetary ephemerides to fit the dynamical modeling of the planet motions. For the three spacecraft, solar corona corrections were not applied in the computation of the spacecraft orbits. Neither the conjunction periods included in the computation of the planetary orbits.

In section 3, we introduce the solar corona modeling and the fitting techniques that were applied to the range bias data. In section 4, the results are presented and discussed. In particular, we discuss the new fitted parameters, the obtained average electron density, and the comparisons with the estimations found in the literature. The impact of these results on planetary ephemerides and new estimates of the asteroid masses are also discussed in this section. The conclusions of this work are given in section 5.



\section{Data analysis of MGS, MEX, and VEX spacecraft}
\subsection{Overview of the MGS mission}
The MGS began its Mars orbit insertion on 12 September 1997. After almost sixteen months of orbit insertion, the aerobraking event converted the elliptical orbit into an almost circular two-hour polar orbit with an average altitude of 378 km. The MGS started its mapping phase in March 1999 and lost communication with the ground station on 2 November 2006. The radio science data collected by the {DSN} consist of one-way Doppler, 2/3 way ramped Doppler and two-way range observations. The radio science instrument used for these data sets consists of an ultra-stable oscillator and the normal MGS transmitter and receiver. The oscillator provides the reference frequency for the radio science experiments and operates on the X-band 7164.624 MHz uplink and 8416.368 MHz downlink frequency. Detailed information of observables and reference frequency are given in \cite{Moyer}.

\subsubsection{ MGS data analysis with GINS}
The radio science data used for  MGS are available on the {NASA} {PDS} Geoscience website\footnote{http://geo.pds.nasa.gov/missions/ mgs/rsraw.html}. These observations were analyzed with the help of the GINS (Géodésie par Intégrations Numériques Simultanées) software provided by the CNES (Centre National d'Etudes Spatiales). GINS numerically integrates the equations of motion and the associated variational equations. It simultaneously retrieves the physical parameters of the force model using an iterative least-squares technique. Gravitational and non-gravitational forces acting on the spacecraft are taken into account. The representation of the MGS spacecraft {\itshape macro-model} and the dynamic modeling of the orbit used in the GINS software are described in \cite{JMarty}. 

For the orbit computation, the simulation was performed by choosing two day data-arcs with two hours (approx. one orbital period of  MGS) of overlapping period. From the overlapping period, we were then able to estimate the quality of the spacecraft orbit determination by taking orbit overlap differences between two successive data-arcs. The least-squares fit was performed on the complete set of Doppler- and range-tracking data-arcs corresponding to the orbital phase of the mission using the DE405 ephemeris \citep{DE405}. To initialize the iteration, the initial position and velocity vectors of MGS were taken from the SPICE {NAIF} kernels\footnote{http://naif.jpl.nasa.gov/naif/}.

The parameters that were estimated during the orbit fitting are (1) the initial position and velocity state vectors of the spacecraft, (2) the scale factors F$_D$ and F$_S$ for drag acceleration and solar radiation pressure, (3) the Doppler- and range residuals per data-arc, (4) the {DSN} station bias per data arc, and (5) the overall range bias per data-arc to account the geometric positions error between the Earth and the Mars. 

\subsubsection{ Results obtained during the orbit computation}
\begin{figure*}
\centering
\includegraphics[width=16cm]{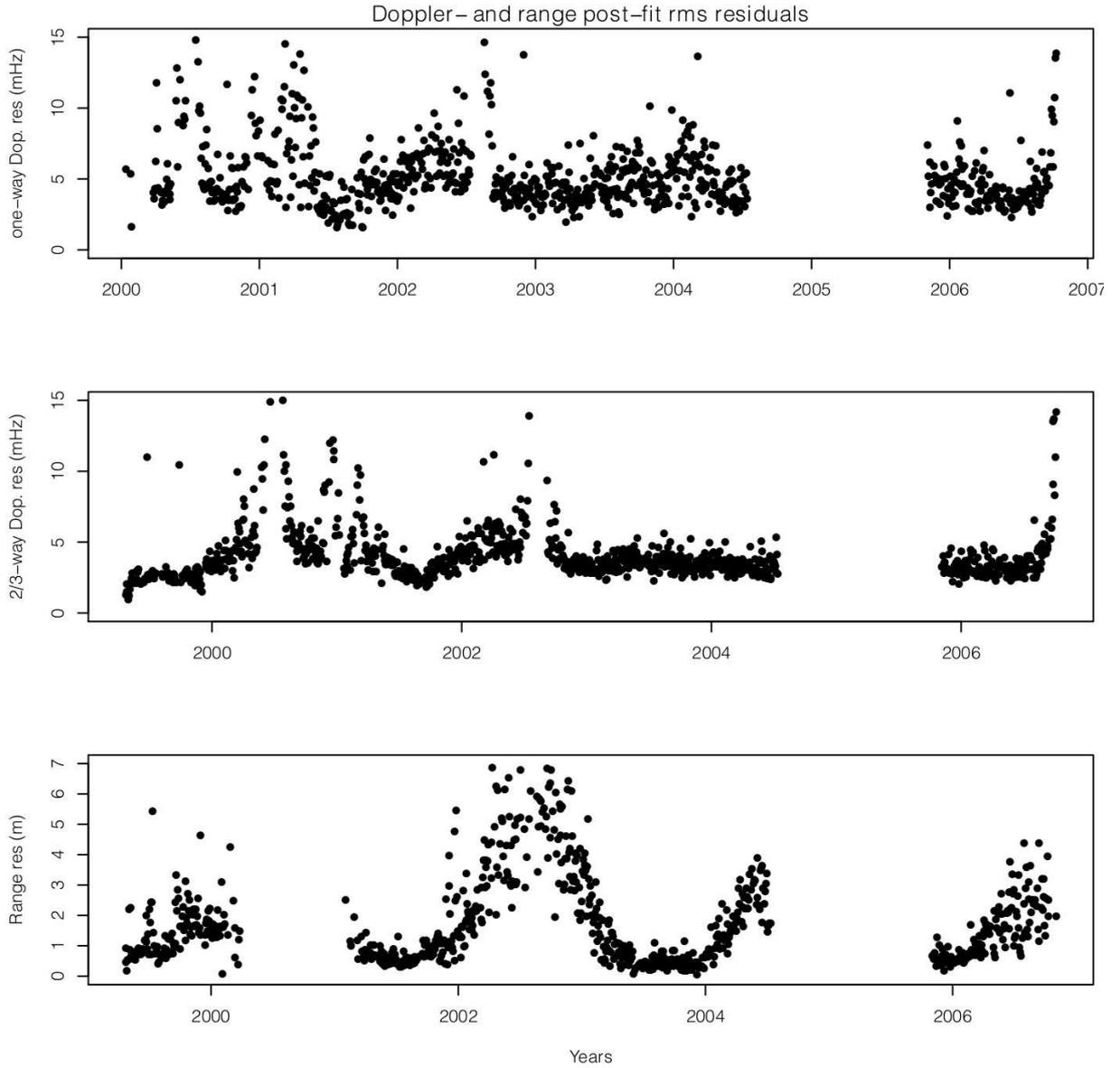} 
\caption{Doppler- and range rms of post-fit residuals of  MGS for each two day data-arc. The residuals show the accuracy of the orbit determinations. The peaks and gaps in residuals correspond to solar conjunction periods of  MGS.}
\label{DOP}
\end{figure*}

In Figure \ref{DOP}, we plot the root mean square (rms) values of the Doppler- and range post-fit residuals estimated for each data-arc. These post-fit residuals represent the accuracy of the orbit determination. To plot realistic points, we did not consider 19$\%$ of the data-arcs during which the rms value of the post-fit Doppler residuals are above 15 mHz, the range residuals are above 7 m, and the drag coefficients and solar radiation pressures have unrealistic values. In Figure \ref{DOP}, the peaks and the gaps in the post-fit residuals correspond to solar conjunction periods. The average value of the post-fit Doppler- and two-way range residuals are less than 5mHz and 1m, which excludes the residuals at the time of solar conjunctions.  


\subsection{MEX and VEX data analysis}
The MEX and VEX radiometric data were analyzed done by the {ESA} navigation team. These data consist of two-way Doppler- and range measurements. These data sets were used for the orbit computations of MEX and VEX. However, despite their insignificant contribution to the accuracy of the orbit computation, range measurements are mainly used for the purpose of analyzing errors in the planetary ephemerides. These computations are performed with the DE405 \citep{DE405} ephemeris. The range bias obtained from these computations is provided by {ESA}, and we compared them to light-time delays computed with the planetary ephemerides ({INPOP}), version 10b, \citep{Fienga2011b}, and the DE421 \citep{DE421} ephemerides. For more details, see \cite{Fienga2009}.

\subsubsection{MEX: orbit and its accuracy}
 Mars Express is the first {ESA} planetary mission for Mars, launched on 2 June 2003. It was inserted into Mars orbit on 25 December 2003. The orbital period of MEX is roughly 6.72 hours and the low polar orbit attitude ranges from 250 km (pericenter) to 11500 km (apocenter).  

The MEX orbit computations were made using 5-7-day track data-arcs with an overlapping period of two days between successive arcs. The differences in the range residuals computed from overlapping periods are less than 3 m, which represents the accuracy of the orbit determination. As described in \cite{Fienga2009}, there are some factors that have limited the MEX orbit determination accuracy, such as the imperfect calibrations of thrusting and the off-loading of the accumulated angular momentum of the reaction wheels and the inaccurate modeling of solar radiation pressure forces.

\begin{figure*}
\centering
\includegraphics[width=16cm]{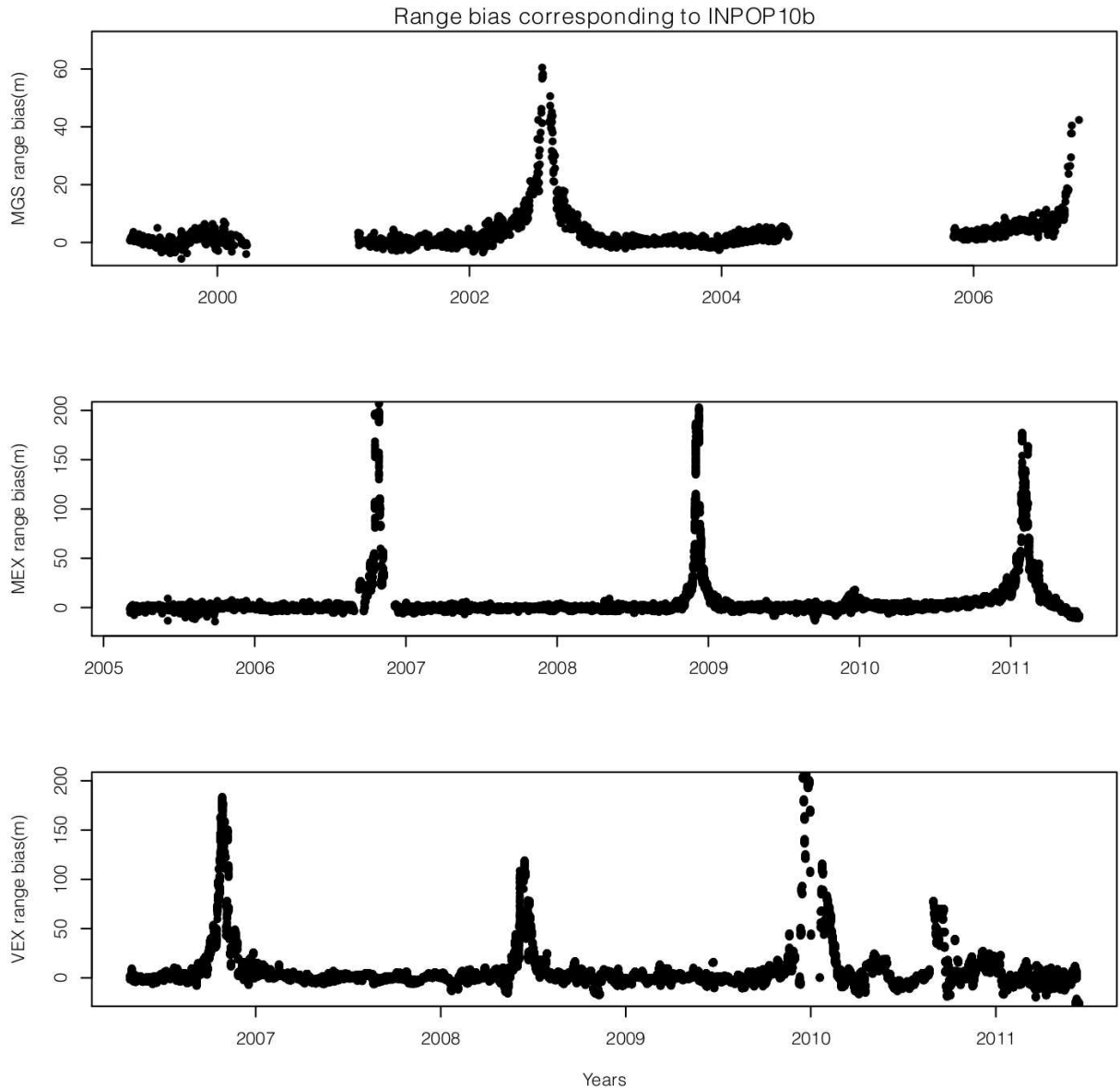} 
\caption{Systematic error (range bias) in the Earth-Mars and the Earth-Venus distances obtained from the INPOP10b ephemeris: (Top panel) range bias corresponding to the MGS obtained for each two day data-arc; (middle and bottom panels) range bias corresponding to the MEX and the VEX. }
\label{MMV}
\end{figure*}

\subsubsection{VEX: orbit and its accuracy}
Venus Express is the first {ESA} planetary mission for Venus, launched on 9 November 2005. It was inserted into Venus orbit on 11 April 2006. The average orbital period of VEX is roughly 24 hours and its highly elliptical polar orbit attitude ranges from 185km (pericenter) to 66500km (apocenter). However, data were almost never acquired during the descending leg of orbit, nor around periapsis \citep{Fienga2009}. 

The orbit was computed in the same manner as the MEX. The computed orbit accuracy is more degraded than MEX \citep{Fienga2009}. This can be explained by the unfavorable patterns of tracking data-arcs, the imperfect calibration of the wheel off-loadings, the inaccurate modeling of solar radiation pressure forces, and the characteristics of the orbit itself. The differences in the range residuals computed from overlapping periods are from a few meters to ten meters, even away from the solar conjunction periods.    

\subsection{Solar conjunction: MGS, MEX, VEX}
The MGS, MEX, and VEX experienced several superior solar conjunctions. In June 2002, when the solar activity was maximum, the MGS {SEP} angle (see Figure \ref{GEO}) remained below 10$^\circ$ for two months and went at minimum to 3.325$^\circ$ according to available data. For MEX, the {SEP} angle remained below 10$^\circ$ for two months and was at minimum three times: in October 2006, December 2008, and February 2011. Similarly, the VEX {SEP} angle remained below 8$^\circ$ for two months and was at minimum in October 2006 and June 2008. The MEX and VEX superior conjunctions happened during solar minima. The influences of the solar plasma on radio signals during solar conjunction periods have been noticed through post-fit range and Doppler residuals, obtained during the orbit computations. Owing to insufficient modeling of the solar corona perturbations within the orbit determination software, no correction was applied during the computations of the spacecraft orbit and range rate residuals. The peaks and gaps shown in Figure \ref{MMV} demonstrate the effect of the solar conjunction on the range bias. The effect of the solar plasma during the MEX and VEX conjunctions on the radiometric data are described in \cite{Fienga2009}, and for the MGS it is shown in Figure \ref{DOP}. The range bias (Figure \ref{MMV}) during solar conjunctions is used for deriving the electron density profiles of a solar corona model. These profiles are derived separately from the orbit determination.     


\section{Solar corona model}
\subsection{Model profile}
\begin{figure}[htbp]
\centering
\includegraphics[width=9cm]{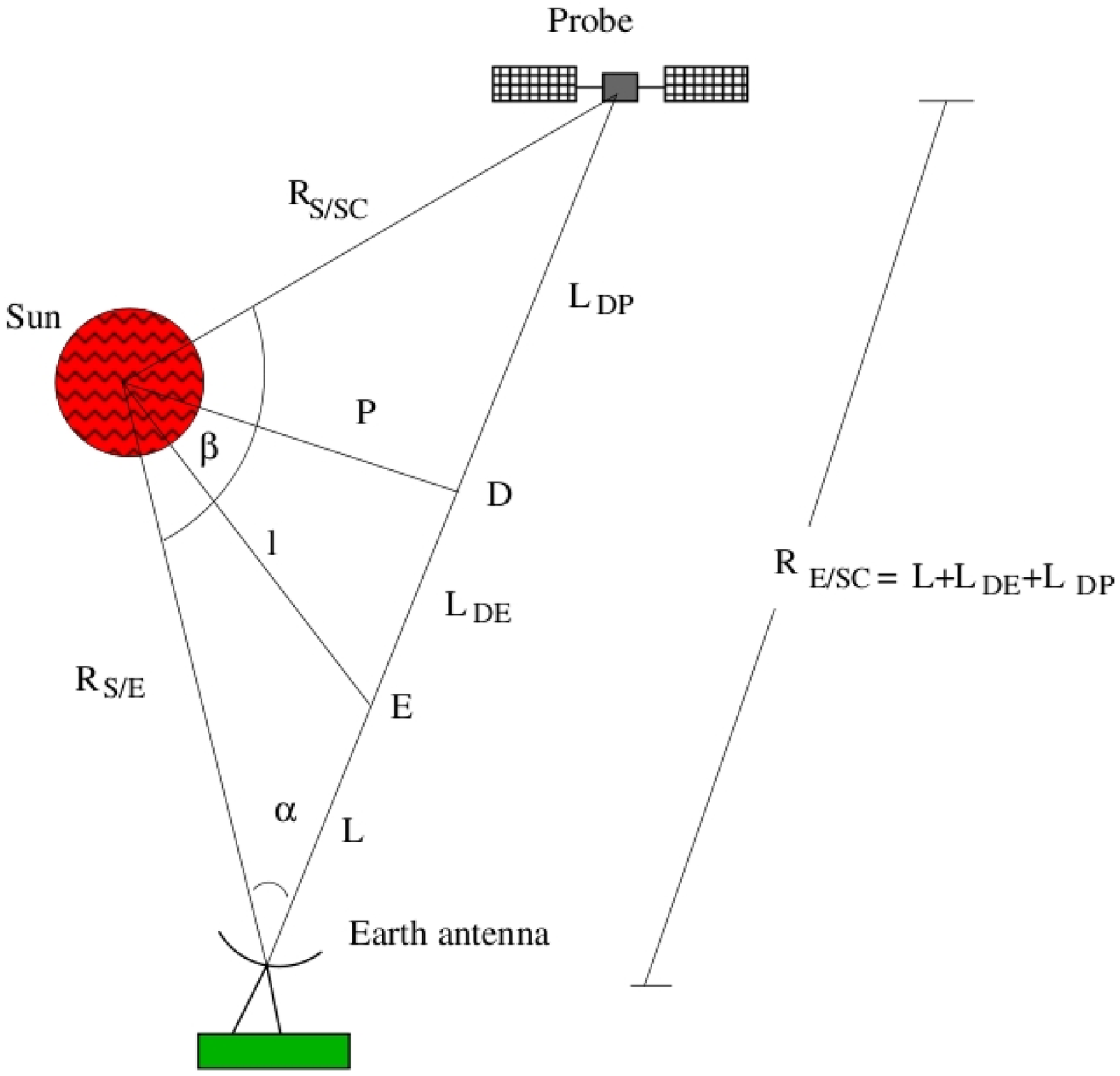}
\caption{Geometric relation between Earth-Sun-Probe. Where $\beta$ is the Earth-Sun-Probe (ESP) angle and $\alpha$ is the Sun-Earth-Probe (SEP) angle.}
\label{GEO}
\end{figure}

As described in section 1, propagations of radio waves through the solar corona cause a travel-time delay between the Earth station and the spacecraft. These time delays can be modeled by integrating the entire ray path (Figure \ref{GEO}) from the Earth station ({\itshape L$_{{Earth}_{s/n}}$}) to the spacecraft ({\itshape L$_{s/c}$}) at a given epoch. This model is defined as

 \begin{equation}
 \Delta \tau = \frac{1}{2cn_{cri} (f)} \times \int_{L_{{Earth}_{s/n}}} ^{L_{s/c}}N_{\mathrm{e}}(l) \ dL
 \end{equation}
 
 \begin{displaymath}
   n_{cri} (f) = 1.240 \times 10^4\ \bigg(\frac{f}{1\ MHz}\bigg)^2 \ \ \mathrm{cm^{-3}}  \ \ ,
 \end{displaymath}
 
 where $c$ is the speed of light, $n_{cri}$ is the critical plasma density for the radio carrier frequency $f$, and $N_{e}$ is an electron density in the unit of electrons per cm$^{3}$ and is expressed as \citep{Bird96}
 
 \begin{equation}
N_e{(l,\theta)} =   B\ \bigg(\frac{l}{R_\odot}\bigg)^{-\epsilon} F(\theta)  \ \ \mathrm{cm^{-3}}  \ \ ,
 \end{equation}

where $B$ and $\epsilon$ are the real positive parameters to be determined from the data. R$_\odot$ and $l$ are the solar radius and radial distance in AU. F($\theta$) is the heliolatitude dependency of the electron density \citep{Bird96}, where $\theta$ represents the heliolatitude location of a point along the {LOS} at a given epoch. The maximum contribution in the electron density occurs when $l$ equals the {MDLOS}, $p$ (see Figure \ref{GEO} ), from the Sun. At a given epoch, {MDLOS} is estimated from the planetary and spacecraft ephemerides. The ratio of the {MDLOS} and the solar radii (R$_\odot$) is given by $r$, which is also called the impact factor:

\begin{displaymath}
 \bigg(\frac{p}{R_\odot}\bigg)=r \ \ .
 \end{displaymath}

 The electron density profile presented by \cite{Bird96} is valid for {MDLOS} greater than 4R$_\odot$. Below this limit, turbulences and irregularities are very high and non-negligible. The solar plasma is therefore considered as inhomogeneous and additional terms (such $r^{-6}$ and $r^{-16}$) could be added to Equation 2 \citep{Muhleman77, Bird94}. However, because of the very high uncertainties in the spacecraft orbit and range bias measurements within these inner regions, we did not include these terms in our solar corona corrections. 
  

At a given epoch, the MGS, MEX, and VEX {MDLOS} always remain in the ecliptic plane. The latitudinal variations in the coverage of data are hence negligible compared to the variation in the {MDLOS}. These data sets are thus less appropriate for the analysis of the electron density as a function of heliolatitude \citep{Bird96}. Equation 2 can therefore be expressed as a function of the single-power-law ($\epsilon$) of radial distance only and be reduced to
\begin{equation}
N_e{(l)} =   B\ \bigg(\frac{l}{R_\odot}\bigg)^{-\epsilon}  \ \ \mathrm{cm^{-3}}  \ \ .
 \end{equation}


On the other hand, at a given interval of the {MDLOS} in the ecliptic plane, \cite{Guhathakurta96} and \cite{Leblanc} added one or more terms to Equation 3, that is

 \begin{equation}
N_e{(l)} =  A\bigg(\frac{l}{R_\odot}\bigg)^{-c} +  B\bigg(\frac{l}{R_\odot}\bigg)^{-d}  \ \ \mathrm{cm^{-3}}  
 \end{equation}
with c $\simeq$ 4 and d = 2.

 To estimate the travel-time delay, we analytically integrated the {LOS} (Equation 1) from the Earth station to the spacecraft, using Equation 3 and Equation 4 individually. The analytical solutions of these integrations are given in Appendix A.

 In general, the parameters of the electron density profiles differ from one model to another. These parameters may vary with the type of observations, with the solar activity, or with the solar wind state. In contrast, the primary difference between the several models postulated for the electron density ($r$ \textgreater 4) is the parameter $\epsilon$ (Equation 3), which can vary from 2.0 to 3.0. 

For example, the density profile parameters obtained by \cite{Muhleman77} using the {\itshape Mariner-7} radio science data for the range of the {MDLOS} from 5R$_\odot$ to 100R$_\odot$ at the time of maximum solar cycle phase are

\begin{displaymath}
N_e =  \frac{(1.3 \pm 0.9) \times 10^{8}}{r^{6}} + \frac{(0.66 \pm 0.53) \times 10^{6}}{r^{2.08 \pm 0.23}}  \ \ \mathrm{cm^{-3}}  \ \ .
 \end{displaymath}

The electron density profile derived  by \cite{Leblanc} using the data obtained by the $Wind$ radio and plasma wave investigation instrument, for the range of the {MDLOS} from about 1.3R$_\odot$ to 215R$_\odot$ at the solar cycle minimum is
 
\begin{displaymath}
N_e =  \frac{0.8 \times 10^{8}} {r^{6}} + \frac{0.41 \times 10^{7}}{r^{4}} + \frac{0.33 \times 10^{6}}{ r^{2}}  \ \ \mathrm{cm^{-3}} \ \ .
 \end{displaymath}
 
Furthermore, based on {\itshape in situ} measurements, such as those obtained with the {\itshape Helios 1 and 2} spacecraft, \cite{Bougeret} gave an electronic profile as a function of the {MDLOS} from 64.5R$_\odot$ to 215R$_\odot$ as follows
 \begin{displaymath}
N_e = \frac{6.14}{p^{2.10}}  \ \ \mathrm{cm^{-3}} \ \ .
 \end{displaymath}        

Similarly, \cite{Issautier98} analyzed {\itshape in situ} measurements of the solar wind electron density as a function of heliolatitude during a solar minimum. The deduced electron density profile at high latitude (\textgreater 40$^\circ$) is given as
\begin{displaymath}
N_e = \frac{2.65}{p^{2.003\pm0.015}}   \ \ \mathrm{cm^{-3}} \ \ .
 \end{displaymath} 
This {\itshape Ulysses} high-latitude data set is a representative sample of the stationary high-speed wind. This offered the opportunity to study the {\itshape in situ} solar wind structure during the minimal variations in the solar activity. 
As presented in \cite{Issautier98}, electronic profiles deduced from other observations in numerous studies were obtained either during different phases of solar activity (minimum or maximum), different solar wind states (fast or slow-wind), using data in the ecliptic plane (low latitudes). These conditions may introduce some bias in the estimation of electronic profiles of density.

Comparisons of the described profiles with the results obtained in this study are presented in section 4.2 and are plotted in Figure 9. 

 
 \subsection{Solar wind identification of {LOS} and data fitting} 

As described in \cite{Schwenn06}, the electronic profiles are very different in slow- and fast-wind regions. In slow-wind regions, one expects a higher electronic density close to the {MNL} of the solar corona magnetic field at low latitudes \citep{You07}. It is then necessary to identify if the region of the {LOS} is either affected by the slow-wind or by the fast-wind. To investigate that question, we computed the projection of the {MDLOS} on the Sun surface. We then located the {MDLOS} heliographic longitudes and latitudes with the maps of the solar corona magnetic field as provided by the {WSO}\footnote{http://wso.stanford.edu/}. This magnetic field is calculated from photospheric field observations with a potential field model\footnote{http://wso.stanford.edu/synsourcel.html}. 

\begin{figure*}
\centering
\includegraphics[width=16cm]{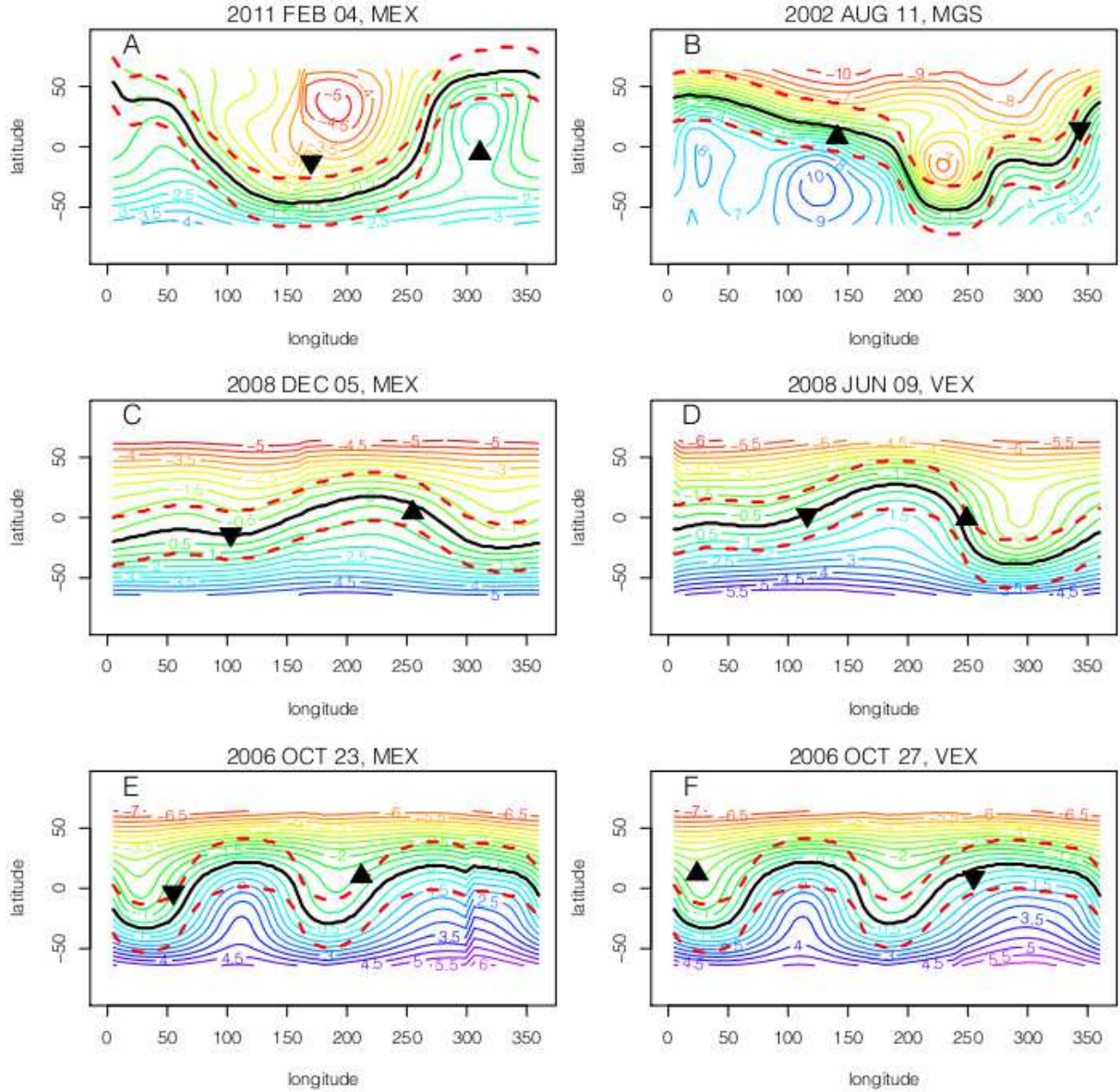} 
\caption{ Solar corona magnetic field maps extracted from {WSO} at the mean epoch of 20R$_\odot$ ingress and egress. The dark solid line represents the magnetic neutral line. The dashed red lines correspond to the  belt of the slow-wind region. The two marked points give the projected locations of the ingress ($\blacktriangle$) and egress ($\blacktriangledown$) minimal distances at 20R$_\odot$. For the {MGS} 2002 solar conjunction, the hypothesis of a $\pm$20$^\circ$ belt is not relevant (see section 4.1.1)}
\label{WIND}
\end{figure*}
However, zones of slow-wind are variable and not precisely determined \citep{Mancuso00,Tokumaru10}. As proposed in \cite{You07,You12}, we took limits of the slow solar wind regions as a belt of 20$^\circ$ above and below the {MNL} during the solar minima. For the 2002 solar maximum, this hypothesis is not valid, the slow-wind region being wider than during solar minima. \cite{Tokumaru10} showed the dominating role of the slow-wind for this entire period and for latitudes lower than $\pm$70$^\circ$ degrees.

An example of the {MDLOS} projection on the Sun$\textquotesingle$s surface with the maps of solar corona magnetic field is shown in Figure \ref{WIND}. 
\begin{figure*}
\centering
\includegraphics[width=16cm]{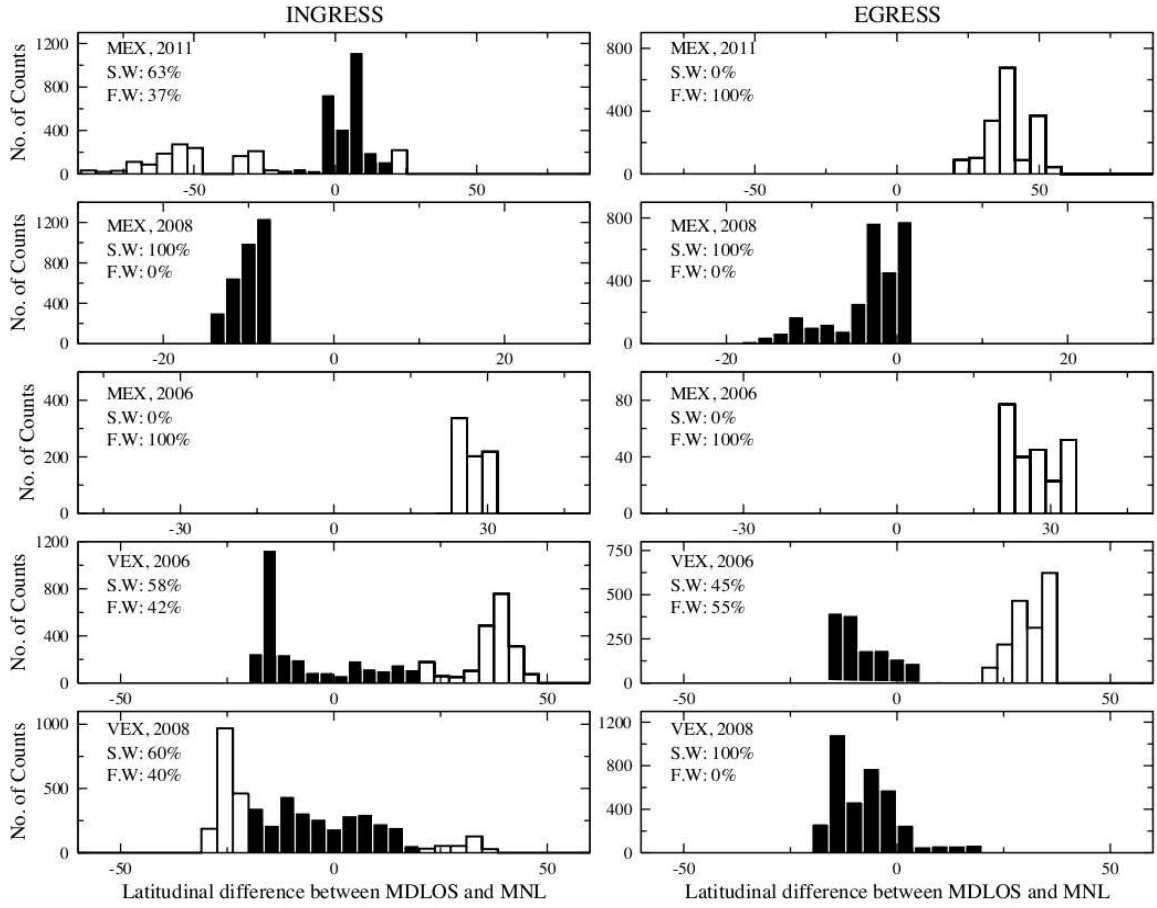} 
\caption{Distribution of the latitudinal differences between {MDLOS} and {MNL} in the slow- and fast-wind regions during the ingress and egress phases of solar conjunctions. The black (white) bars present the slow- (fast-) wind regions as defined by $\pm$20$^\circ$ ($\textgreater$$\pm$20$^\circ$) along the {MNL} during solar minima.}
\label{histogram}
\end{figure*}
 \begin{figure*}
\centering
 \includegraphics[width=15cm]{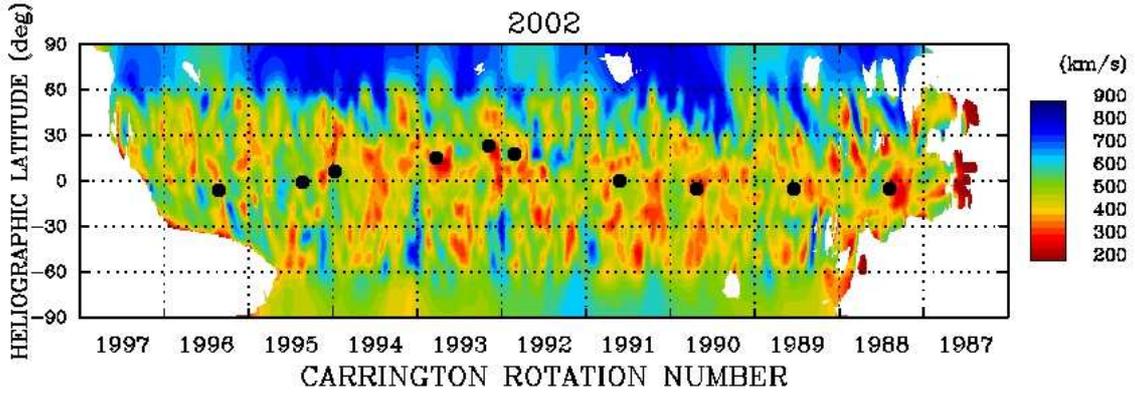} 
\caption{2002 synoptic source surface map of solar wind speeds derived from STEL IPS observations extracted from \cite{Tokumaru10}. The black dots represent the {MGS} {MDLOS} during the 2002 conjunction period for the range of 12R$_\odot$ to 120R$_\odot$.}
\label{IPS}
\end{figure*}
These magnetic field maps correspond to the mean epoch of the ingress and egress phases of the solar conjunction at 20R$_\odot$. The dark solid line represents the {MNL} and the belt of the slow-wind region is presented by dashed lines. The two marked points give the projected locations of the {MDLOS}, ingress ($\blacktriangle$) and egress ($\blacktriangledown$) at 20R$_\odot$. For the entire {MDLOS}, this distribution of the slow- or fast-wind regions in the ingress and egress phases of solar conjunctions are shown in Figure \ref{histogram}. The black (white) bars in Figure \ref{histogram} present the count of data sets distributed in the slow- (fast-) wind region. 


\begin{sidewaystable*}[p]
\caption{Solar corona model parameters and electron densities estimated from two different models using the MGS, MEX, and VEX range bias. The electron density out from the interval of the {MDLOS} is extrapolated from the given model parameters.}
\centering
\renewcommand{\arraystretch}{2.0}
\small
\begin{threeparttable}
\begin{tabular}{ccccccccccccccc}\Xhline{2\arrayrulewidth}

\multicolumn{1}{c}{\multirow{2}{*}{{\bf S/C}}}    &\multicolumn{1}{c}{\multirow{2}{*}{{\bf Year}}}    &\multicolumn{1}{c}{\multirow{2}{*}{{\bf S.C\tnote{1}}}}    &\multicolumn{1}{c}{\multirow{2}{*}{{\bf S.W.S\tnote{2}}}}   &\multicolumn{1}{c}{\multirow{2}{*}{{\bf {MDLOS}\tnote{3}}}}  &\multicolumn{6}{c}{{ \bf $N_e$} {\bf $=$ B$r^{-\epsilon}$}}    & \multicolumn{4}{c}{{\bf $N_e$} {\bf $=$A$r^{-4}$ {\bf  $+$  B$r^{-2}$}}} \\
\cline {7-10} \cline {7-10} \cline{12-15} \cline{12-15}
&& &&&&{\bf B (}$\times$ {\bf $10^6$)} &{\bf $\epsilon$ }& {\bf Ne @ $20R_\odot$} & {\bf Ne @ $215R_\odot$\tnote{*}} &&{\bf B (}$\times${\bf $10^6$)}&{\bf A (}$\times$ {\bf $10^8$)}&{\bf Ne @ $20R_\odot$} & {\bf Ne @ $215R_\odot$\tnote{*}}\\ \Xhline{2\arrayrulewidth}
  
\multicolumn{1}{c}{\multirow{2}{*}{MGS}} & \multicolumn{1}{c}{\multirow{2}{*}{Aug 2002}} &\multicolumn{1}{c}{\multirow{2}{*}{Max}}& S.W&12-215  &&0.51$\pm$0.06 & 2.00$\pm$0.01&1275$\pm$150&11$\pm$1.5&&
0.52$\pm$0.04 & 0&1300$\pm$100&11$\pm$1     \\ 
\multicolumn{1}{c}{}                        &
\multicolumn{1}{c}{} &  & F.W & - & &- & -&-&- && - & -&-&-    \\  \hline

\multicolumn{1}{c}{\multirow{2}{*}{MEX}} & \multicolumn{1}{c}{\multirow{2}{*}{Oct 2006}} &\multicolumn{1}{c}{\multirow{2}{*}{Min}}& S.W 
& - & &- & -&-&- && - & -&-&-  \\ 
\multicolumn{1}{c}{}                        &
\multicolumn{1}{c}{} &  & F.W &6-40 & &1.90$\pm$0.50 & 2.54$\pm$0.07&942$\pm$189&2.3$\pm$0.9&& 0.30$\pm$0.10&0.16$\pm$0.02&850$\pm$263&6$\pm$1  \\  \hline

\multicolumn{1}{c}{\multirow{2}{*}{MEX}} & \multicolumn{1}{c}{\multirow{2}{*}{Dec 2008}} &\multicolumn{1}{c}{\multirow{2}{*}{Min}}&S.W & 6-71&&
 0.89$\pm$0.42& 2.40$\pm$0.16& 673$\pm$72&2.3$\pm$1.3 & & 
0.22$\pm$0.04&0.10$\pm$0.02&615$\pm$90&5$\pm$1  \\ 
\multicolumn{1}{c}{}                        &
\multicolumn{1}{c}{} &  & F.W & - & &- & -&-&- && - & -&-&-  \\  \hline

\multicolumn{1}{c}{\multirow{2}{*}{MEX}} & \multicolumn{1}{c}{\multirow{2}{*}{Feb 2011}} &\multicolumn{1}{c}{\multirow{2}{*}{Min}}&S.W &40-152 && 0.52$\pm$0.10& 2\tnote{**}&1300$\pm$25\tnote{*}&11$\pm$1  & &
0.52$\pm$0.10&0&1300$\pm$25\tnote{*}&11$\pm$1  \\ 
\multicolumn{1}{c}{}                        &
\multicolumn{1}{c}{} &  & F.W& 6-60  && 1.70$\pm$0.10& 2.44$\pm$0.01&1138$\pm$31&3.5$\pm$0.5 && 0.33$\pm$0.02&0.24$\pm$0.04&975$\pm$60&7$\pm$1  \\   \Xhline{2\arrayrulewidth}

\multicolumn{1}{c}{\multirow{2}{*}{VEX}} & \multicolumn{1}{c}{\multirow{2}{*}{Oct 2006}} &\multicolumn{1}{c}{\multirow{2}{*}{Min}}&S.W &12-154 && 0.52$\pm$0.30 & 2.10$\pm$0.10&964$\pm$600&7$\pm$4&&
0.40$\pm$0.28&0&1000$\pm$600&9$\pm$6  \\ 
\multicolumn{1}{c}{}                        &
\multicolumn{1}{c}{} &  & F.W &12-130 && 1.35$\pm$1.10& 2.33$\pm$0.30&1256$\pm$649&5$\pm$2 && 0.44$\pm$0.15 & 0&1087$\pm$480&9$\pm$4  \\  \hline

\multicolumn{1}{c}{\multirow{2}{*}{VEX}} & \multicolumn{1}{c}{\multirow{2}{*}{Jun 2008}} &\multicolumn{1}{c}{\multirow{2}{*}{Min}}& S.W&12-154 && 1.70$\pm$1.50 & 2.50$\pm$0.50&950$\pm$625&3$\pm$2  &&
0.31$\pm$0.20&0&775$\pm$450&7$\pm$4  \\ 
\multicolumn{1}{c}{}                        &
\multicolumn{1}{c}{} &  & F.W& 41-96 && 0.10$\pm$0.01 & 2\tnote{**}&250$\pm$25\tnote{*}&2$\pm$1&&0.10$\pm$0.01 & 0&250$\pm$25\tnote{*}&2$\pm$1  \\  \hline

 \end{tabular}
\begin{tablenotes}
\item[1] S.C: Solar cycle
\item[2] S.W.S: Solar wind state (F.W: fast-wind,    S.W: slow-wind)
\item[3] MDLOS: Minimum distance of the line of sight in the unit of solar radii (R$_\odot$)
\item[*] Extrapolated value
\item[**] Fixed 
\end{tablenotes}
\end{threeparttable}
\label{MMVTABLE}
\end{sidewaystable*}


After separating the {MDLOS} into slow- and fast-wind regions as defined in Figure \ref{histogram}, the parameters of Equations 3 and 4 are then calculated using least-squares techniques. These parameters are obtained for various ranges of the {MDLOS}, from 12R$_\odot$ to 215R$_\odot$ for  MGS, 6R$_\odot$ to 152R$_\odot$ for MEX, and from 12R$_\odot$ to 154R$_\odot$ for VEX. The adjustments were performed, for all available data acquired at the time of the solar conjunctions, for each spacecraft individually, and separately for fast- and slow-wind regions (see Table \ref{MMVTABLE}).

To estimate the robustness of the electronic profile determinations, adjustments on ingress and egress phases were performed separately. The differences between these two estimations and the one obtained on the whole data set give the sensitivity of the profile fit to the distribution of the data, but also to the solar wind states. These differences are thus taken as the uncertainty in the estimations and are given as error bars in the Table \ref{MMVTABLE}.



\section{Results and discussions}
\subsection{Estimated model parameters and electron density}

As described in section 3.2, we estimated the model parameters and the electron density separately for each conjunction of the MGS, MEX, and VEX. A summary of these results is presented in Table \ref{MMVTABLE}. The {MDLOS} in the unit of solar radii (R$_\odot$) mentioned in this table (column 5) represents the interval of available data used for calculating the electronic profiles of density. These profiles were then used for extrapolating the average electron density at 215R$_\odot$ (1AU). The period of the solar conjunctions, solar activities, and solar wind states are also given in columns 2, 3, and 4. Table \ref{MMVTABLE} also contains the estimated parameters of two different models: the first model from \cite{Bird96} corresponds to Equation 3, whereas the second model is based on \cite{Guhathakurta96} and \cite{Leblanc} and follows Equation 4 with c=4. Estimated model parameters for the slow- and fast-wind regions are presented in columns 6 and 7.

\subsubsection{Mars superior conjunction}
\begin{figure*}
\centering
\includegraphics[width=14cm]{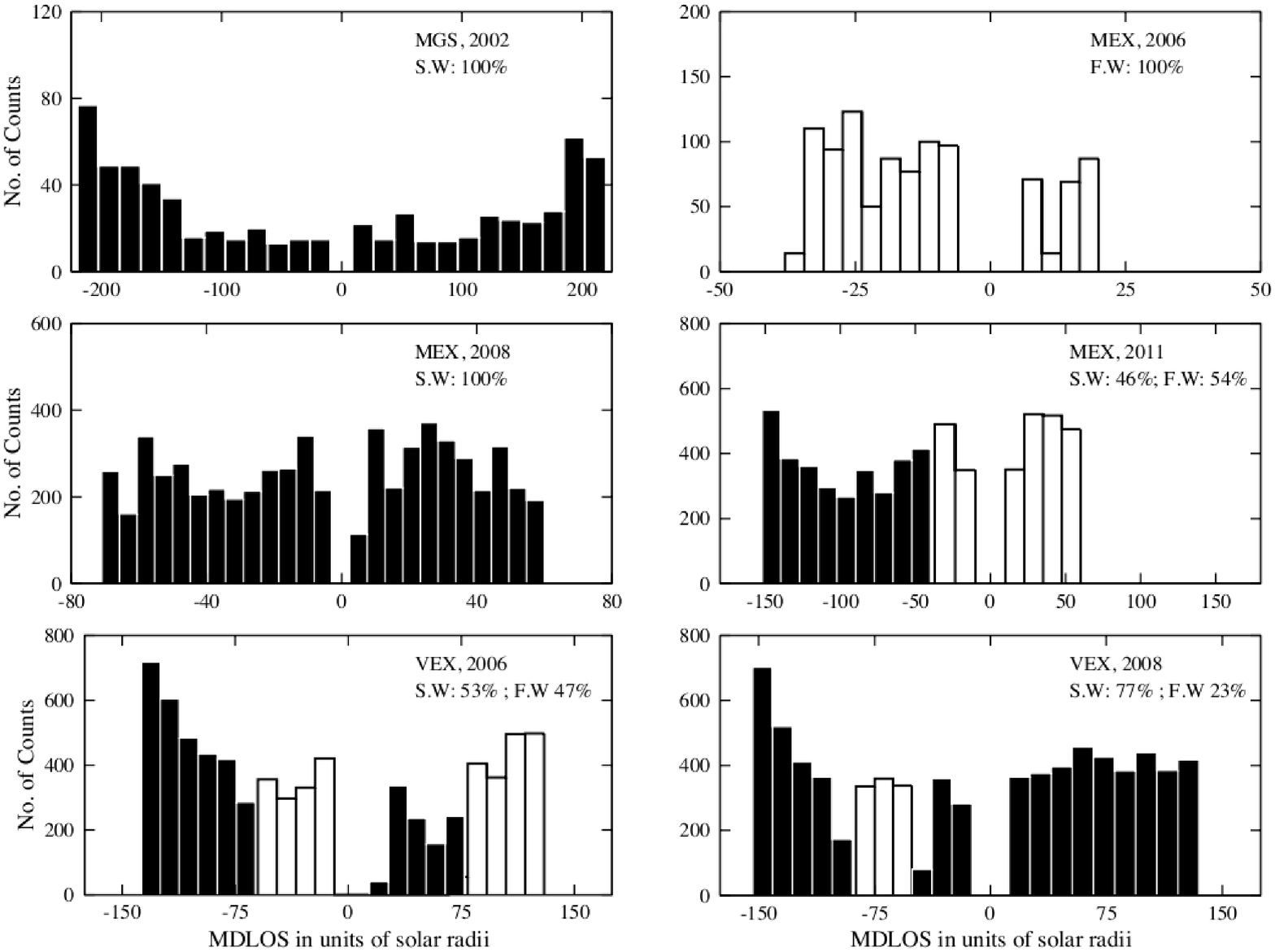} 
\caption{Distribution of the MGS, MEX, and VEX data in the slow (black) and fast (white) wind with respect to {MDLOS} in units of solar radii (R$_\odot$). Negative (positive) {MDLOS} represents the distribution in the ingress (egress) phase.}
\label{BARPLOT}
\end{figure*}

The MGS experienced its superior conjunction in 2002 when the solar activity was maximum and the slow-wind region was spread at about $\pm$70$^\circ$ of heliolatitude \citep{Tokumaru10}. Figure \ref{IPS} represents the projection of the MGS {MDLOS} on the solar surface (black dots) superimposed with the 2002 synoptic source surface map of solar wind speeds derived from STEL IPS observations extracted from \cite{Tokumaru10}. It suggests that, the {MDLOS} of the {MGS} exclusively remains in the slow-wind region. Respective estimates of the model parameters and of the electron densities are given in Table \ref{MMVTABLE}. From this table one can see that the estimates of the electron density from both models are very similar. Parameter $\epsilon$ of Equation 3 is then estimated as 2.00$\pm$0.01, which represents a radially symmetrical behavior of the solar wind and hence validates the assumption of a spherically symmetrical behavior of the slow-wind during solar maxima \citep{Guhathakurta96}. Whereas for Equation 4, the contribution of the $r^{-4}$ term at large heliocentric distances (r $\textgreater$ 12R$_\odot$) is negligible compared to the $r^{-2}$ term. Thus, the parameter $A$ for these large heliocentric distances is fixed to zero and consequently gave similar results to Equation 3, as shown in Table \ref{MMVTABLE}.

\begin{table*}
\caption{Electron densities estimated from different models at 20R$_\odot$ and at 215R$_\odot$ (1AU). }
\centering
\renewcommand{\arraystretch}{1.4}
\small
\begin{threeparttable}
\begin{tabular}{ c c c c c c}
 \hline  
                        &                           &                                            \\
    Authors & Spacecraft  &    Solar  &{MDLOS} & Ne @ 20R$_\odot$ & Ne @ 215R$_\odot$  \\ 
     &                         &          activity&&      (el. cm$^-3$)& (el. cm$^-3$)                      \\

 \hline
\cite{Leblanc} & {\itshape Wind}&  Min&1.3-215 &847&  7.2 \\

\cite{Bougeret}  & {\itshape Helios 1 and 2}& Min/Max & 65-215&  890& 6.14\\
    
 \cite{Issautier98} & {\itshape Ulysses} & Min & 327-497 & 307\tnote{*} & 2.65$\pm$0.5\tnote{*}\\

\cite{Muhleman77} &{\itshape Mariner} \itshape6 $and$ \itshape7 &Max. &5-100&1231$\pm$ 64&9$\pm$3\\
     
\cite{Bird94} & {\itshape Ulysses} &  Max &  5-42& 1700$\pm$ 100 & 4.7$\pm$0.415 \\
 \cite{AndersonV2}&{\itshape Voyager}  2 &  Max & 10-88& 6650$\pm$ 850 & 38$\pm$4\\
    \hline
    
\end{tabular}
\begin{tablenotes}
\item[*] Mean electron density corresponds to latitude $\geq$40$^\circ$ 
\end{tablenotes}
\end{threeparttable}
\label{comparison}
\end{table*}

\begin{figure}[htbp]
\centering
\includegraphics[width=9cm]{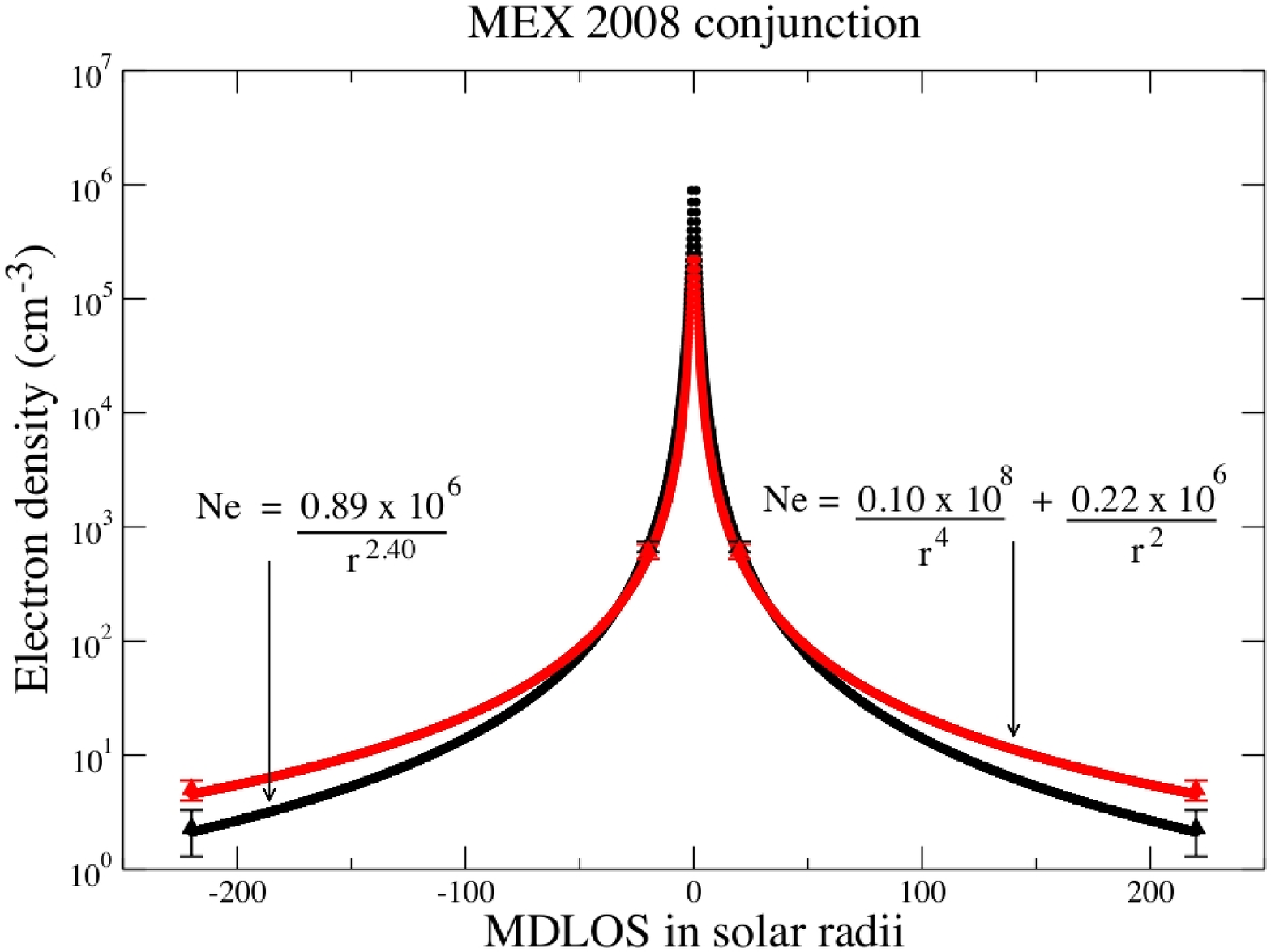} 
\caption{Example of the comparison between electron density models given in Equations 3 and 4. The electron density profiles are plotted from 1R$_\odot$ to 215R$_\odot$ (1AU) for the MEX 2008 conjunction using the model parameters given in Table \ref{MMVTABLE}. The error bars plotted in the figure correspond to the electron density obtained at 20R$_\odot$ and 215R$_\odot$ (see Table \ref{MMVTABLE}).}
\label{M1VSM2}
\end{figure} 

In contrast, MEX experienced its superior conjunctions in 2006, 2008, and 2011 during solar minima. The distribution of data during these conjunctions with respect to {MDLOS} are shown in Figure \ref{BARPLOT}. From this figure, one can see that the MEX 2006 (2008) conjunction corresponds to the fast- (slow-) wind region, whereas 2011 conjunction is a mixture of slow- and fast-winds. The estimated parameters of these conjunctions are given in Table \ref{MMVTABLE}. An example of the comparison between two models (Equation 3 and 4) is shown in Figure \ref{M1VSM2}. This figure compares the electron density profiles obtained from the two models during the MEX 2008 conjunction. These profiles are extrapolated from 1R$_\odot$ to 6R$_\odot$ and from 71R$_\odot$ to 215R$_\odot$. The  upper triangles in Figure \ref{M1VSM2} indicate electronic densities obtained at 20R$_\odot$ and at the extrapolated distance of 215R$_\odot$ (1AU) with error bars obtained as described in section 3.2. As one sees in that figure, electronic profiles are quite similar over the computation interval till 71R$_\odot$ and become significantly different after this limit.
This suggests that the estimated parameters for both models are valid for the range of {MDLOS} given in Table \ref{MMVTABLE}. 
Finally, for the MEX 2011 conjunction, as shown in Figures \ref{histogram} and \ref{BARPLOT}, the data are mainly distributed in the slow-wind (63\%) during ingress phase and in the fast-wind (100\%) during egress phase. Owing to the unavailability of slow-wind data near the Sun ({MDLOS} $\textless$ 40R$_\odot$), we fixed $\epsilon$ to 2 for Equation 3 (see Table \ref{MMVTABLE}). Moreover, the average electron density estimated for this conjunction is higher for the slow-wind than for the fast-wind and it is consistent with \cite{Tokumaru10}, which suggests that near the {MNL}, the electron content is higher than in the fast-wind regions.

\subsubsection{Venus superior conjunction}
The VEX 2006 and 2008 conjunctions exhibit a mixture of slow- and fast-wind (Figure \ref{BARPLOT}). These conjunctions occurred approximately at the same time as the MEX superior conjunctions. However, the limitations in the VEX orbit determination introduced bias in the estimation of the model parameters and the electron densities. This can be verified from the discrepancies presented in Table \ref{MMVTABLE} for the VEX 2006 and 2008 conjunctions. Despite these high uncertainties, post-fit range bias corrected for the solar corona allows one to add complementary data in the construction of the planetary ephemerides (see section 4.5). 


\subsection{Comparison with other models}
\begin{figure*}
\centering
\includegraphics[width=14cm]{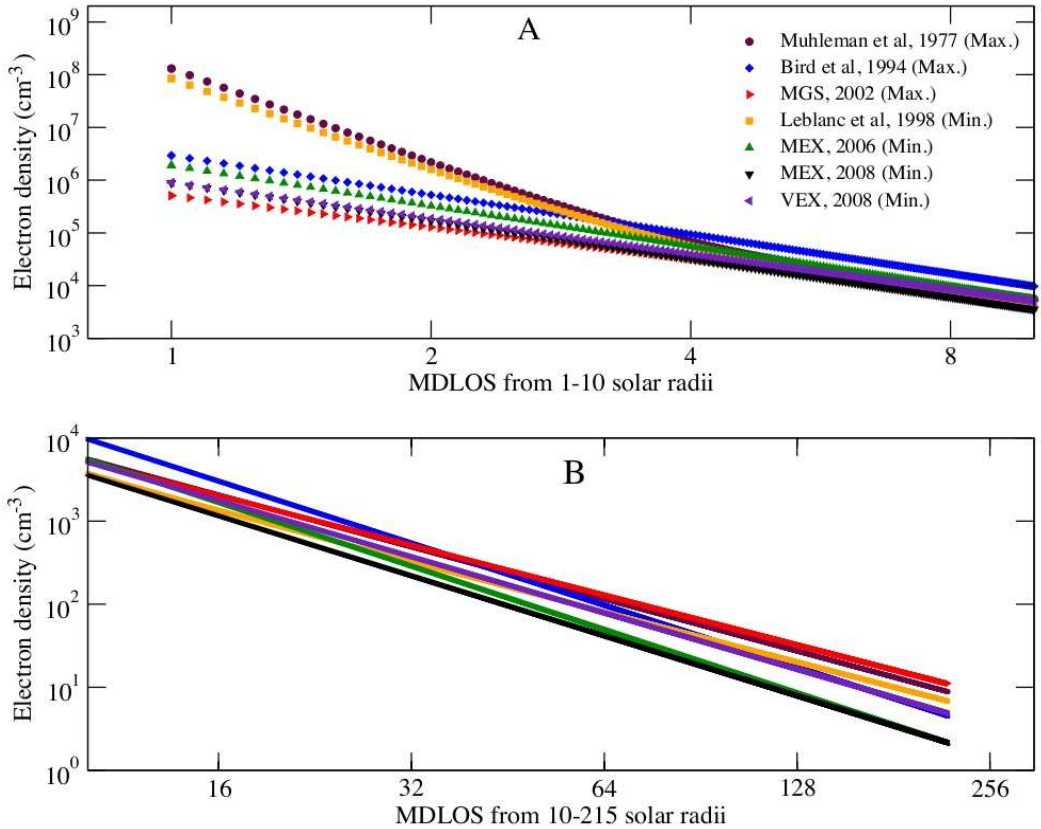}  
\caption{Comparison of different electron density profiles at different phases of the solar cycle from 1R$_\odot$ to 215R$_\odot$ (1AU).}
\label{LOG}
\end{figure*} 
Table \ref{comparison} represents the estimated electron densities at 20R$_\odot$ and 215R$_\odot$ (1AU) from the various models. These models are representative of radio science measurements \citep{Muhleman77,AndersonV2,Bird94,Bird96}, {\itshape in situ} measurements \citep{Bougeret,Issautier97}, and solar type III radio emission \citep{Leblanc} measurements (Table \ref{literature}). Table \ref{comparison} and Figure \ref{LOG} allow us to compare the average electron density, obtained from different observations, made approximately during the same solar activity cycle.

Figure \ref{LOG} illustrates the comparisons of different electron density profiles, extrapolated from 1R$_\odot$ to 215R$_\odot$. From this figure it can be seen that approximately all electron density profiles follow similar trends ($\propto$  $r^{-\epsilon}$, $\epsilon$ varying from 2 to 3) until 10R$_\odot$ (panel $B$), whereas the dispersions in the profiles below 10R$_\odot$ (panel $A$) are due to the contribution of higher order terms, such as $r^{-4}$, $r^{-6}$ or $r^{-16}$.

In Table \ref{comparison}, we also provide the average electron density at 20R$_\odot$ and 215R$_\odot$, based on the corresponding models (if not given by the authors). The two individual electron density profiles for ingress and egress phases have been given by \cite{AndersonV2} and \cite{Bird94}. To compare their estimates with ours, we took the mean values of both phases. Similarly, \cite{Muhleman77} gave the mean electron density at 215R$_\odot$ estimated from round-trip propagation time delays of the {\itshape Mariner 6 and 7} spacecraft.
 
Table \ref{comparison} shows a wide range of the average electron densities, estimated at 20R$_\odot$ and 215R$_\odot$ during different phases of solar activity. Our estimates of the average electron density shown in Table \ref{MMVTABLE} are very close to the previous estimates, especially during solar minimum. The widest variations between our results and the earlier estimates were found during solar maxima and can be explained from the high variability of the solar corona during these periods.

\subsection{Post-fit residuals}
\begin{figure*}[tbp]
\centering
\includegraphics[width=16cm]{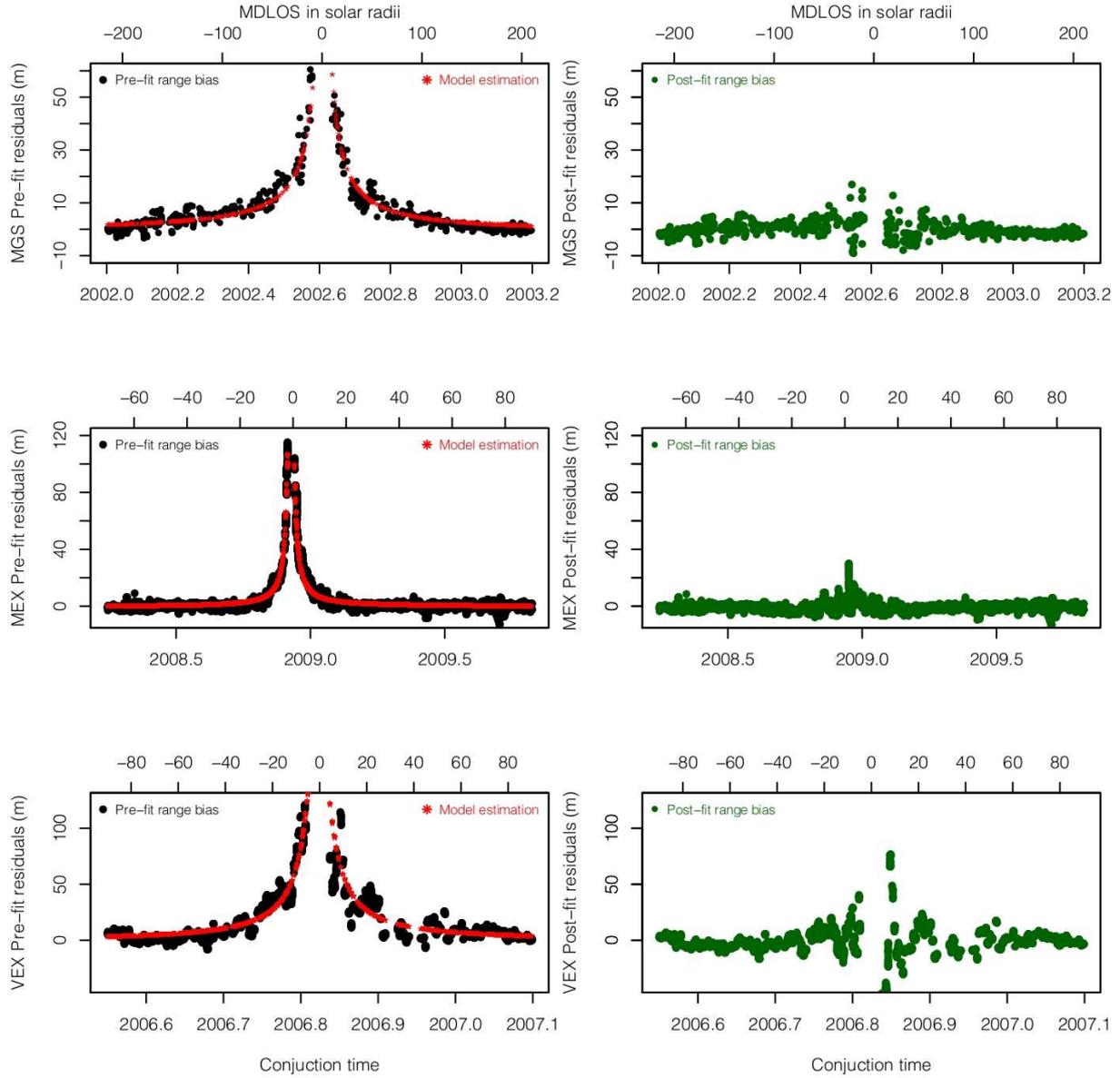}  
\caption{Left panel: model-estimated solar corona over-plotted on the pre-fit residuals. Right panel: post-fit residuals after corona corrections. Top, middle, and bottom panels correspond to the MGS 2002, MEX 2008, and VEX 2006 conjunctions.}
\label{POSTFIT}
\end{figure*} 
One of the objectives of this study is to minimize the effect of the solar corona on the range bias. These post-fit range biases can then be used to improves the planetary ephemeris ({INPOP}). The pre-fit range bias represents the systematic error in the planetary ephemerides during the solar conjunction periods. Figure \ref{POSTFIT} shows the pre-fit residuals over plotted with the simulated time delay (in units of distance), obtained from the solar corona model based on Equation 3. In contrast, the post-fit range bias represents the error in the ephemerides after correction for solar corona perturbations. 

\begin{table}
\caption{Statistics of the range bias before and after solar corona corrections.}
\centering
\renewcommand{\arraystretch}{1.4}
\small
\begin{threeparttable}
\begin{tabular}{ccccccc}\Xhline{2\arrayrulewidth}

\multicolumn{1}{c}{\multirow{2}{*}{{\bf S/C}}} &\multicolumn{4}{c}{{\bf Pre-fit}}     & \multicolumn{2}{c}{{\bf Post-fit}} \\ \cline{3-4} \cline{6-7}
&& {\bf mean (m)} &{\bf $\sigma$ (m) }&&{\bf mean (m)} &{\bf $\sigma$ (m)} \\ \Xhline{2\arrayrulewidth}

MGS, 2002&& {6.02} &{10.10}&&{-0.16} &{2.89} \\ 
MEX, 2006&& {42.03} &{39.30}&&{0.85} &{9.06} \\
MEX, 2008&& {16.00} &{20.35}&&{-0.10} &{4.28} \\
MEX, 2011&& {15.44} &{19.20}&&{0.11} &{6.48} \\
VEX, 2006&& {5.47} &{11.48}&&{-0.74} &{6.72} \\
VEX, 2008&& {3.48} &{11.48}&&{-0.87} &{7.97} \\
\Xhline{2\arrayrulewidth}
\end{tabular}
\end{threeparttable}
\label{STAT}
\end{table}

From Figure \ref{POSTFIT} one can see that the systematic trend of the solar corona perturbations is almost removed from the range bias. The post-fit range bias of the VEX at low solar radii (especially during the egress phase) is not as good as the MGS and MEX. This can be explained by the degraded quality of the VEX orbit determination (see section 2.2.2). The dispersion in the pre-fit and post-fit range bias is given in Table \ref{STAT}. The estimated dispersions in the post-fit range bias are one order of magnitude lower than the dispersions in the pre-fit range bias. It shows a good agreement between the model estimates and the radiometric data. The corrected range bias (post-fit) is then used to improve the planetary ephemerides (see section 4.5). 

\subsection{Model parameter dependency on the ephemerides.} 
The range bias data are usually very important for constructing the planetary ephemerides \citep{DE421,Fienga2011b}. These measurements correspond to at least 57\% of the total amount of data used for the {INPOP} construction \citep{Fienga2009}. Range bias data at the time of the solar conjunctions are not taken into account due to very high uncertainties (see Figure \ref{MMV}). Equations 3 and Figure \ref{GEO} show the dependency of the density profile over geometric positions of the spacecraft (orbiting a planet) relative to the Earth and the Sun. The range bias used for this study includes the error in the geometric distance of Mars and Venus relative to the Earth. These errors are varying from one ephemeris to another and impact directly on the estimates of the mean electron density.    

\begin{figure*}[tbp]
\centering
\includegraphics[width=12cm]{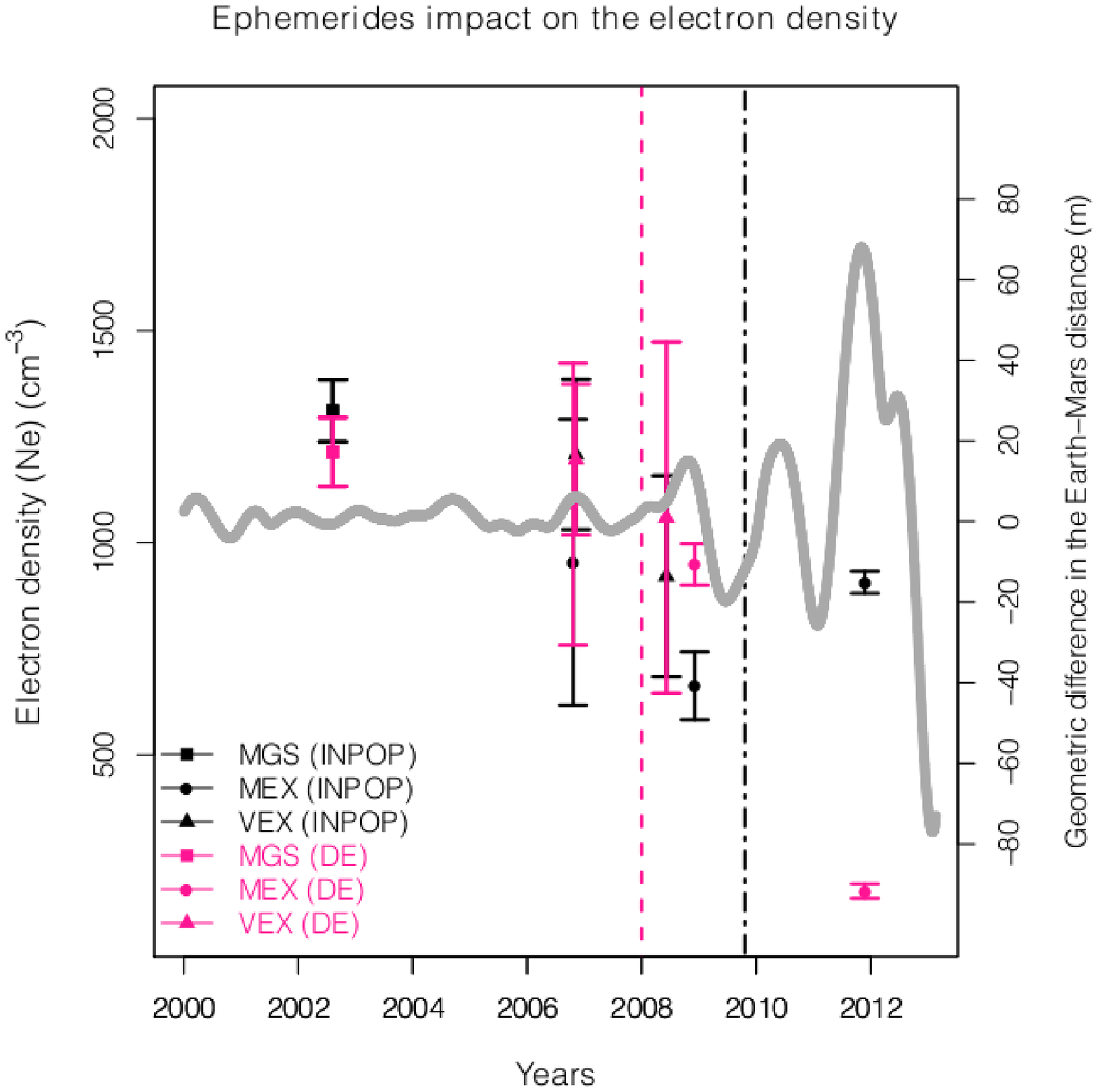}  
\caption{Variation of the average electron density at 20R$_\odot$ and 215R$_\odot$ using the DE421 and INPOP10b ephemeris. The dotted-dashed ({INPOP}) and dashed (DE421) vertical lines present the starting time of extrapolation. The plain line shows the differences in the Mars-Earth geometric distances estimated with INPOP10b and DE421}
\label{EPH}
\end{figure*} 

Figure \ref{EPH} illustrates the electron density estimated at 20R$_\odot$ for MGS, {MEX} and {VEX} using DE421 and INPOP10b ephemerides. The dashed (DE421) and dotted-dashed (INPOP10b) lines present the time limit up to which these ephemerides are fitted over range bias data. The curve represents the differences between INPOP10b and DE421 estimations of Mars-Earth geometric distances. 

In particular, the Mars orbit is affected by the belt of asteroids. The asteroid masses may cause a degradation in the estimates of the Mars orbit. Therefore, as one can see in Figure \ref{EPH}, the geometric differences between the INPOP10b and DE421 estimates of the Mars-Earth geometric distances are magnified from the extrapolation period onward. Hence, the electron densities estimated using DE421and INPOP10b are consistent with each other within the error bars before the extrapolation period, whereas after this period, the DE421and INPOP10b estimates of the electron densities are quite different from each other. The sensitivity of the solar corona parameters and the electron densities, deduced from the analysis of the range bias, is then low as long as the computation is included in the time interval of the fit of the planetary ephemerides. However, out from the fitting time, the quality of these computations can be degraded by the extrapolation capability of the planetary ephemerides. Conversely, by fitting the planetary ephemerides  ({INPOP}) including data corrected for the solar corona perturbations, some noticeable improvement can appear in the extrapolation capability of the planetary ephemerides and in the estimates of the asteroid masses (see section 4.5).

\subsection{Impact on planetary ephemerides.}
As one can see in Figure \ref{POSTFIT}, correct the effects induced by the solar corona on the observed Mars-Earth distances is significant over some specific periods of time (during solar conjunctions). We aim to estimate the impact of this important but time-limited improvement of the measurements of interplanetary distances on the construction of the planetary ephemerides.
To evaluate any possible improvement, we produced two ephemerides, INPOP10c and INPOP10d, both fitted over the same data set as was used for the construction of INPOP10b \citep{Fienga2011b}. This data set contains all planetary observations commonly used for {INPOP} (see \cite{Fienga2009, Fienga2011}), including the MGS data obtained in section 2.1.1 and the MEX and VEX range bias provided by {ESA}. These newly built ephemerides are based on the same dynamical modeling as described in  \cite{Fienga2009, Fienga2011}. However, INPOP10c is estimated without any solar corona corrections on the MGS, MEX and VEX range bias, and INPOP10d includes the solar corona corrections evaluated in the previous sections.
The selection of the fitted asteroid masses and the adjustment method (bounded value least-squares associated with a priori sigmas) are the same for  the two cases. The weighting schema are also identical. The differences remain in the quality and the quantity of the range bias used for the fit (one corrected for solar plasma and one not) and in the procedure selecting the data actually used in the fit.

For INPOP10c, about 119901 observations were selected. Of these, 57$\%$ are MGS, MEX, and VEX range bias data that are not corrected for solar corona effects. Based on a very conservative procedure, observations obtained two months before and after the conjunctions were removed from the fitted data sample. This strategy leads to removal of about 7$\%$ of the whole data set, which represents 14$\%$ of the MGS, MEX, and VEX observations.
For INPOP10d, thanks to the solar corona corrections, only observations of {SEP} smaller than 1.8 degrees were removed from the data sample. This represents less than 1$\%$ of the whole data sample. The estimated accuracy of the measurements corrected for the solar plasma is 2.4 meters when observations not affected by the solar conjunctions have an accuracy of about 1.7 meters. By keeping more observations during solar conjunction intervals, the number of data with a good accuracy is then significantly increased.

Adjustments of planet initial conditions, mass of the sun, sun oblateness, mass of an asteroid ring, and the masses of 289 asteroids were then performed in the same fitting conditions as INPOP10b.

No significant differences were noted for the evaluated parameters except for the asteroid masses.


For the masses estimated both in INPOP10c and INPOP10d, 20$\%$ induce perturbations bigger than 5 meters on the Earth-Mars distances during the observation period. The masses of these 59 objects are presented in Tables \ref{masscompar1}, \ref{masscompar2}, and \ref{masscompar2b}. Within the 1-$\sigma$ uncertainties deduced from the fit, we notice 10 (17$\%$) significant differences in masses obtained with INPOP10c and INPOP10d, quoted with a  \textquotedblleft *$\textquotedblright$ in column 5 of the two tables, 7 (12$\%$) new estimates made with INPOP10d, noted N in column 5, and 6 (10$\%$) masses put to 0 in INPOP10d when estimated in INPOP10c, marked with 0 in the fifth column. 

Table \ref{massconnues} lists masses found in the literature compared with the quoted values of Tables \ref{masscompar1}, \ref{masscompar2}, and \ref{masscompar2b}. In this table, 80$\%$ of the INPOP10d estimates have a better consistency with the values obtained by close encounters than the one obtained with INPOP10c. Of these, the new estimates obtained with INPOP10d for (20), (139) and (27) agree well with the values found in the literature. For (45) Eugenia, the INPOP10c value is closer to the mass deduced from the motion of its satellite \citep{2008Icar..196...97M} even if the INPOP10d value is still compatible at 2-$\sigma$. For (130) Elektra, the INPOP10c and INPOP10d estimated values are certainly under evaluated. 
Finally, one can note the systematic bigger uncertainties of the INPOP10d estimates. The supplementary data sample collected during the solar conjunctions that has 30$\%$ more noise than the data collected beyond the conjunction can explain the  degradation of the uncertainties for the INPOP10d determinations compared to INPOP10c. 

By correcting the range bias for the solar corona effects, we added more informations related to the perturbations induced by the asteroids during the conjunction intervals.   

In principal, during the least-squares estimation of the asteroid masses, the general trend of the gravitational perturbation induced by the asteroid on the planet orbits should be described the most completely by the observable (the Earth-Mars distances) without any lack of information. In particular, for an optimized estimation, the data sets used for the fit should include local maxima of the perturbation. 

However, it could happen that some of the local maxima occur during the solar conjunction intervals. One can then expect a degradation of the least-squares estimation of the perturber mass if no solar corrections are applied or if these intervals are not taken into account during the fit. To estimate which mass determination can be more degraded than another by this {\it window} effect, we estimated $L$, the percentage of local maxima rejected from the INPOP10c fit in comparison with the INPOP10d adjustment including all data sets corrected for solar plasma. 
$L$ will give the loss of information induced by the rejection of the solar conjunction intervals in terms of highest perturbations.

The $L$ criteria are given in column 7 of Tables \ref{masscompar1}, \ref{masscompar2}, and \ref{masscompar2b}. As an example, for (24) Themis one notes in Table \ref{massconnues} the good agreement between the close encounter estimates and the INPOP10d mass determination compared with INPOP10c. On the other hand, based on the $L$ criteria, 36$\%$ of the local maxima happen near solar conjunctions.
By neglecting the solar conjunction intervals, more than a third of the biggest perturbations are missing in the adjustment. This can explain the more realistic INPOP10d estimates compared with INPOP10c. 

We also indicate in Tables \ref{masscompar1}, \ref{masscompar2}, and \ref{masscompar2b} if important constraints were added in the fit (column 8). In these cases, even if new observations are added to the fit (during the solar conjunction periods), there is a high probability to obtain a stable estimates of the constrained masses as for the biggest perturbers of Table \ref{masscompar1}. For the other mass determinations, one can note a consistency between high values of the $L$ criteria and the non-negligible mass differences between INPOP10c and INPOP10d. 

 \begin{figure}[htbp]
\centering
\includegraphics[width=8cm]{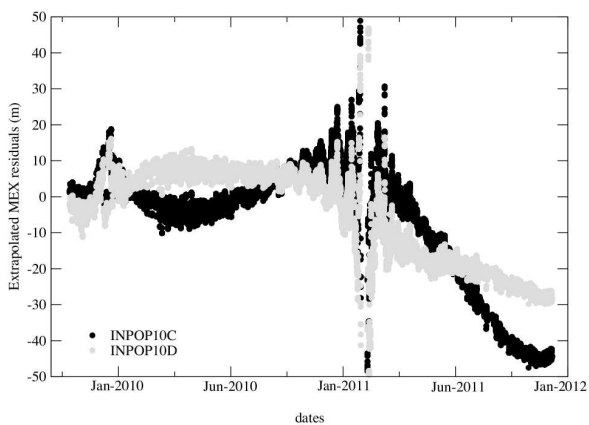}  
\caption{MEX extrapolated residuals estimated with INPOP10d (light dots) and INPOP10c (dark dots).}
\label{extpol}
\end{figure}

By improving the range bias residuals during the solar conjunction periods, we then slightly improved the asteroid mass determinations. 
 
Estimates of residuals for data samples not used in the {INPOP} fit and dated after or before the end of the fitting interval are currently made to evaluate the real accuracy of the planetary ephemerides \citep{Fienga2011,Fienga2009}. To estimate if the use of the solar corona corrections induces a global improvement of the planetary ephemerides, the MEX extrapolated residuals were computed with INPOP10c and INPOP10d. As one can see in Figure \ref{extpol}, the INPOP10d MEX extrapolated residuals show  a better long-term behavior compared with INPOP10c with 30$\%$ less degraded residuals after two years of extrapolation. 

Supplementary data of the MEX and VEX obtained during the first six months of 2012 would confirm the long-term evolutions of the INPOP10d, INPOP10c and INPOP10b. 

This improvement can be explained by the more realistic adjustment of the ephemerides with denser data sets (7$\%$) and more consistent asteroid mass fitting.

\begin{table*}[!p]
\caption{ Masses of the 59 asteroids that induce perturbations greater than 5 meters on the Earth-Mars distances during the period of observations  \citep{Kuchynka2010}. Columns 2 and 3 give the values of the masses with the 1-$\sigma$ uncertainties obtained with INPOP10c and INPOP10d. The differences between INPOP10c and INPOP10d values are given in column 4. Column 5 indicates if the INPOP10d masses are newly determined (quoted N), set equal to 0 when estimated by INPOP10c (quoted 0), significantly different from INPOP10c determinations (*). The maximum impacts of the perturbations are given in column 5. In column 7 is given the L criteria, the rate of local maxima rejected from the INPOP10c fit but included in the INPOP10d. In the last column we provide the status of the imposed constraints: $S$ for strong and $N$ for normal.}
\centering
\renewcommand{\arraystretch}{1.4}
\small
\begin{tabular}{c l l c c c c c}
\hline
IAU designation	& INPOP10c	& INPOP10d 	  & Diff & S & Impact & L & \\
number	& $10^{12}$ x M$_{\odot}$	& $10^{12}$ x M$_{\odot}$	& $10^{12}$ x M$_{\odot}$ & & m & $\%$ & \\
\hline
     4 & 130.109 $\pm$ 0.716 & 130.109 $\pm$ 0.983 & 0.000 & &1198.953& 20.5 & S\\
    1 & 467.267 $\pm$ 2.047 & 467.267 $\pm$ 2.437 & 0.000 & &793.741& 17.3 & S\\
    2 & 103.843 $\pm$ 1.689 & 102.654 $\pm$ 1.933 & 1.189 & &146.270& 11.8 & S\\
  324 & 5.723 $\pm$ 0.531 & 5.723 $\pm$ 0.611 & 0.000 & &93.536& 1.0 & S\\
   10 & 43.513 $\pm$ 3.300 & 43.513 $\pm$ 3.877 & 0.000 & &77.003& 15.9 & S\\
   19 & 3.884 $\pm$ 0.447 & 3.450 $\pm$ 0.526 & 0.435 & &59.069& 13.8 & N\\
    3 & 11.793 $\pm$ 0.714 & 11.793 $\pm$ 0.803 & 0.000 & &55.639& 0.6 & S\\
  704 & 19.217 $\pm$ 2.315 & 19.217 $\pm$ 2.869 & 0.000 & &34.492& 7.4 & S\\
  532 & 2.895 $\pm$ 1.043 & 2.895 $\pm$ 1.093 & 0.000 & &32.714& 2.3 & S\\
    9 & 3.864 $\pm$ 0.613 & 3.063 $\pm$ 0.665 & 0.801 & &29.606& 20.6 & N\\
    7 & 5.671 $\pm$ 0.512 & 5.367 $\pm$ 0.591 & 0.305 & &27.822& 13.9 & S\\
   29 & 7.629 $\pm$ 1.067 & 7.227 $\pm$ 1.225 & 0.402 & &26.673& 2.9 & S\\
   24 & 7.641 $\pm$ 1.596 & 2.194 $\pm$ 1.775 & 5.447 & * &26.131& 36.0& N\\
   31 & 3.256 $\pm$ 2.034 & 4.411 $\pm$ 2.050 & 1.155 & &23.466& 24.1 & S\\
   15 & 13.576 $\pm$ 0.939 & 13.576 $\pm$ 1.264 & 0.000 & &21.555& 20.6 & S\\
    6 & 7.084 $\pm$ 0.822 & 7.084 $\pm$ 1.048 & 0.000 & &21.150& 7.4 & S\\
   11 & 3.771 $\pm$ 0.976 & 3.771 $\pm$ 1.110 & 0.000 & &17.301& 31.9 & S\\
  139 & 0.000 $\pm$ 0.000 & 3.579 $\pm$ 0.595 & 3.579 & N &16.687& 32.0 & N\\
  747 & 4.129 $\pm$ 0.841 & 6.805 $\pm$ 1.089 & 2.676 & * &15.937& 31.6 &N\\
  105 & 3.111 $\pm$ 0.556 & 3.111 $\pm$ 0.745 & 0.000 & &15.196& 4.5 & N\\
   20 & 0.000 $\pm$ 0.000 & 1.921 $\pm$ 0.661 & 1.921 & N &14.763& 39.9 & N\\
  372 & 12.365 $\pm$ 2.676 & 12.365 $\pm$ 2.990 & 0.000 & &13.796& 19.0 & S\\
    \hline
   \end{tabular}
\label{masscompar1}
\end{table*}

\begin{table*}[bp]
\caption{Same as Table \ref{masscompar1}}
\centering
\renewcommand{\arraystretch}{1.4}
\small
\begin{tabular}{c c c c c c c c}
\hline
IAU designation	& INPOP10c	& INPOP10d 	  & Diff & S& Impact& L\\
number	& $10^{12}$ x M$_{\odot}$	& $10^{12}$ x M$_{\odot}$	& $10^{12}$ x M$_{\odot}$ & & m& $\%$ \\ \hline
   8 & 3.165 $\pm$ 0.353 & 3.325 $\pm$ 0.365 & 0.159 & &12.664& 17.7 & S\\
   45 & 3.523 $\pm$ 0.819 & 1.518 $\pm$ 0.962 & 2.005 & * &11.790& 21.0 & N\\
   41 & 3.836 $\pm$ 0.721 & 2.773 $\pm$ 0.977 & 1.063 & &11.568& 15.2 & N\\
  405 & 0.005 $\pm$ 0.003 & 0.006 $\pm$ 0.003 & 0.001 & &11.378& 21.2 & N\\
  511 & 9.125 $\pm$ 2.796 & 9.125 $\pm$ 3.138 & 0.000 & &10.248& 20.5 & S\\
   52 & 8.990 $\pm$ 2.781 & 8.990 $\pm$ 3.231 & 0.000 & &9.841& 3.0 & S\\
   16 & 12.613 $\pm$ 2.286 & 12.613 $\pm$ 2.746 & 0.000 & &9.701& 8.9 & S\\
  419 & 1.185 $\pm$ 0.461 & 0.425 $\pm$ 0.398 & 0.760 & &9.585& 10.2 & N\\
   78 & 0.026 $\pm$ 0.016 & 0.024 $\pm$ 0.016 & 0.002 & &9.389& 9.8 & N \\
  259 & 0.092 $\pm$ 0.002 & 0.006 $\pm$ 0.003 & 0.086 & &9.222& 31.4 & N\\
   27 & 0.000 $\pm$ 0.000 & 1.511 $\pm$ 0.982 & 1.511 & N &9.146& 29.5 & N\\
 23 & 0.000 $\pm$ 0.000 & 0.093 $\pm$ 0.156 & 0.093 & N &9.067& 31.1 & N\\
  488 & 3.338 $\pm$ 1.850 & 0.000 $\pm$ 0.000 & 3.338 & 0 &8.614& 2.8& N\\
  230 & 0.000 $\pm$ 0.000 & 0.263 $\pm$ 0.169 & 0.263 & N &7.620& 27.0 & N \\
  409 & 0.002 $\pm$ 0.001 & 0.002 $\pm$ 0.001 & 0.000 & 0 &7.574& 2.2 & N \\
   94 & 1.572 $\pm$ 1.097 & 7.631 $\pm$ 2.488 & 6.058 & * &7.466& 28.5 & N\\
  344 & 2.701 $\pm$ 0.497 & 2.088 $\pm$ 0.515 & 0.613 & &7.465& 15.3 & N\\
  130 & 0.099 $\pm$ 0.047 & 0.221 $\pm$ 0.069 & 0.122 & * &7.054& 31.7& N\\
  111 & 1.002 $\pm$ 0.323 & 0.000 $\pm$ 0.000 & 1.002 & 0 &6.985& 11.4& N\\
  109 & 0.495 $\pm$ 0.322 & 1.318 $\pm$ 0.852 & 0.823 & &6.865& 18.7& N\\
   42 & 1.144 $\pm$ 0.362 & 0.083 $\pm$ 0.389 & 1.061 & * &6.829& 0.6& N\\
   63 & 0.000 $\pm$ 0.000 & 0.424 $\pm$ 0.143 & 0.424 & N &6.451& 17.4& N\\
   12 & 2.297 $\pm$ 0.319 & 1.505 $\pm$ 0.331 & 0.792 & * &6.159& 21.7& N\\
  469 & 0.088 $\pm$ 0.073 & 0.000 $\pm$ 0.000 & 0.088 & 0 &6.107& 18.1& N\\
  144 & 0.176 $\pm$ 0.297 & 0.751 $\pm$ 0.361 & 0.575 & &6.087& 22.8& N\\
    \hline
\end{tabular}
\label{masscompar2}
\end{table*}

\begin{table*}[bp]
\caption{Same as Table \ref{masscompar1}}
\centering
\renewcommand{\arraystretch}{1.4}
\small
\begin{tabular}{c c c c c c c c}
\hline
IAU designation	& INPOP10c	& INPOP10d 	  & Diff & S& Impact& L\\
number	& $10^{12}$ x M$_{\odot}$	& $10^{12}$ x M$_{\odot}$	& $10^{12}$ x M$_{\odot}$ & & m& $\%$ \\ \hline
  356 & 4.173 $\pm$ 0.868 & 4.173 $\pm$ 0.902 & 0.000 & &5.759& 2.2& N\\
  712 & 0.000 $\pm$ 0.000 & 1.228 $\pm$ 0.267 & 1.228 & N &5.745& 2.2& N\\
   88 & 1.340 $\pm$ 0.866 & 0.000 $\pm$ 0.000 & 1.340 & 0 &5.742& 1.4& N\\
   60 & 0.402 $\pm$ 0.221 & 0.282 $\pm$ 0.268 & 0.120 & &5.733& 3.8& N\\
   50 & 0.686 $\pm$ 0.187 & 1.031 $\pm$ 0.566 & 0.345 & &5.702& 2.0& N\\
  128 & 4.699 $\pm$ 1.522 & 0.000 $\pm$ 0.000 & 4.677 & 0 &5.624& 3.0& N\\
    5 & 0.448 $\pm$ 0.165 & 0.913 $\pm$ 0.220 & 0.466 & * &5.533& 15.8 & N\\
   59 & 4.332 $\pm$ 0.607 & 1.364 $\pm$ 1.097 & 2.968 & * &5.325& 12.1& N\\
   98 & 1.414 $\pm$ 0.603 & 2.100 $\pm$ 0.705 & 0.686 & &5.195& 15.1& N\\
  194 & 6.387 $\pm$ 0.701 & 4.380 $\pm$ 0.819 & 2.007 & * &5.145& 2.9& N\\
   51 & 3.546 $\pm$ 0.748 & 3.639 $\pm$ 0.937 & 0.093 & &5.109& 15.6 & N\\
  156 & 3.263 $\pm$ 0.438 & 3.089 $\pm$ 0.576 & 0.174 & &5.103& 19.3& N\\
  \hline
\end{tabular}
\label{masscompar2b}
\end{table*}
\begin{table*}[bp]
\caption{Asteroid masses found in the recent literature compared with the values estimated in INPOP10c and INPOP10d.  The uncertainties are given at 1 published sigma.}
\centering
\renewcommand{\arraystretch}{1.4}
\small
\begin{tabular}{c c c c c c}
\hline
IAU designation	& INPOP10c	& Close-encounters	& Refs	& 	INPOP10d  & \\
number	& $10^{12}$ x M$_{\odot}$	& $10^{12}$ x M$_{\odot}$	& 	& $10^{12}$ x M$_{\odot}$ & $\%$	\\ \hline
5 & 0.448 $\pm$ 0.165 & 1.705 $\pm$ 0.348 & \cite{zielenbach2011} & 0.913 $\pm$ 0.220 & 15.8 \\
12 & 2.297 $\pm$ 0.319 & 2.256  $\pm$ 1.910 & \cite{zielenbach2011}  & 1.505 $\pm$ 0.331 & 21.7\\
20 & 0.000 $\pm$ 0.000 & 1.680 $\pm$ 0.350 & \cite{2011Icar..212..438B} &  1.921 $\pm$ 0.661 & 39.9\\
24 & 7.6 $\pm$ 1.6& 2.639 $\pm$ 1.117 & \cite{zielenbach2011}	& 	 2.2 $\pm$ 1.7  & 36.0\\
27 & 0.000 $\pm$ 0.000 & 1.104 $\pm$ 0.732 & \cite{zielenbach2011}  &1.511 $\pm$ 0.982 & 29.5 \\
45 & 3.523 $\pm$ 0.819 & 2.860 $\pm$ 0.060 & \cite{2008Icar..196...97M} &  1.518 $\pm$ 0.962 & 21.0 \\
59 & 4.332 $\pm$ 0.607 & 1.448 $\pm$ 0.0187 & \cite{zielenbach2011}  & 1.364 $\pm$ 1.097 & 12.1 \\ 
94 & 1.572 $\pm$ 1.097 & 7.878 $\pm$ 4.016 & \cite{zielenbach2011}  &7.631 $\pm$ 2.488 & 28.5\\
130 & 0.099 $\pm$ 0.047 & 3.320 $\pm$ 0.200 & \cite{2008Icar..196...97M} &  0.221 $\pm$ 0.069 & 31.7 \\
139 & 0.000 $\pm$ 0.000 & 3.953 $\pm$ 2.429 & \cite{zielenbach2011}  & 3.579 $\pm$ 0.595 & 32.0\\
\hline
\end{tabular}
\label{massconnues}
\end{table*}



\section{Conclusion}
We analyzed the navigation data of the MGS, MEX, and VEX spacecraft acquired during solar conjunction periods. We estimated new characteristics of solar corona models and electron densities at different phases of solar activity (maximum and minimum) and at different solar wind states (slow and fast). Good agreement was found between the solar corona model estimates and the radiometric data. We compared our estimates of electron densities with earlier results obtained with different methods. These estimates were found to be consistent during the same solar activities. During solar minima, the electron densities obtained by {\itshape in situ} measurements and solar radio burst III are within the error bars of the MEX and VEX estimates. However, during the solar maxima, electron densities obtained with different methods or different spacecraft show weaker consistencies. These discrepancies need to be investigated in more detail, which requires a deeper analysis of data acquired at the time of solar maxima.

The {MGS}, {MEX}, and {VEX} solar conjunctions data allows us to analyze the large-scale structure of the corona electron density. These analyses provide individual electron density profiles for slow- and fast-wind regions during solar maxima and minima activities.

In the future, planetary missions such as MESSENGER will also provide an opportunity to analyze the radio-science data, especially at the time of maximum solar cycle.

We tested the variability caused by the planetary ephemerides on the electron density parameters deduced from the analysis of the range bias. This variability is smaller than the 1-$\sigma$ uncertainties of the time-fitting interval of the planetary ephemerides but becomes wider beyond this interval.
Furthermore, data corrected for solar corona perturbations were used for constructing of the {INPOP} ephemerides. Thanks to these supplementary data, an improvement in the estimation of the asteroid masses and a better behavior of the ephemerides were achieved.

\section{Acknowledgments}
A. K. Verma is the research fellow of CNES and Region Franche-Comté and thanks CNES and Region Franche-Comté for financial support. Part of this work was made using the GINS software; we would like to acknowledge CNES, who provided us access to this software. We are also grateful to J.C Marty (CNES) and P. Rosenbatt (Royal Observatory of Belgium) for their support in handling the GINS software. The Authors are grateful to the anonymous referee for helpful comments, which improved the manuscript.
\label{}

\bibliographystyle{aa}
\bibliography{arXive.bib}

\clearpage
\begin{appendix} 
\section{Analytical solution}

This section presents the analytical solutions of Equation 1. Let, $I_{1}$ and $I_{2}$ be the integral solutions of Equation 1 using Equation 3 and 4 $i.e$. ,

 \begin{equation}
    I_{1} = \int_{L_{{Earth}_{s/n}}} ^{L_{s/c}} B\bigg(\frac{l}{R_\odot}\bigg)^{-\epsilon} \ dL 
 \end{equation}
 
 and
    \begin{equation}
   I_{2} = \int_{L_{{Earth}_{s/n}}} ^{L_{s/c}} \bigg[ A\bigg(\frac{l}{R_\odot}\bigg)^{-4}+B\bigg(\frac{l}{R_\odot}\bigg)^{-2} \bigg] \ dL \ \ .
    \end{equation}

From the geometry (Figure 3) we define

   \begin{displaymath}
   P = R_{S/E} \ sin \ \alpha \ \ ,
   \end{displaymath}
   \begin{displaymath}
   L_{DC} = R_{E/SC} - R_{S/E} \ cos \ \alpha \ \ ,
   \end{displaymath}
   \begin{displaymath}
   L_{DE}^2 = l^2 - P^2 \ \ ,
   \end{displaymath}
    \begin{displaymath}
   l^2 = L^2 + R_{S/E}^2 - 2 \ L \ R_{S/E} \ cos \ \alpha \ \ ,
   \end{displaymath}

where $\alpha$ and $\beta$ are the angle between the Sun-Earth-Probe (SEP) and the Earth-Sun-Probe (ESP). $P$ is the {MDLOS} from the Sun. With these expressions, $I_{1}$ can be written as
    \begin{displaymath}
    I_{1} = \int_{L_{{Earth}_{s/n}}} ^{L_{s/c}} B \ R_\odot^{\epsilon} \ \bigg(\frac{dL}{(L^2 + R_{S/E}^2 - 2 \ L \ R_{S/E} \ cos \ \alpha)^{\epsilon/2}}\bigg)  s \end{displaymath}
   
   \begin{displaymath}
  \ \ \ = B \ R_\odot^{\epsilon}  \int_{L_{{Earth}_{s/n}}} ^{L_{s/c}} \ \bigg(\frac{dL}{([L - R_{S/E} \ cos \ \alpha]^2 + \ R_{S/E}^2 \ sin^2 \ \alpha)^{\epsilon/2}}\bigg) \ \ .
   \end{displaymath}
   
   Assuming,
   \begin{displaymath}
   x = L - R_{S/E} \ cos \ \alpha \ \ ,
   \end{displaymath}
   \begin{displaymath}
   a = R_{S/E} \ sin \ \alpha \ \ ,
   \end{displaymath}
   \begin{displaymath}
   dx = dL \ \ ,
   \end{displaymath}

   with {\itshape L} = {\itshape 0} at the Earth station ({\itshape L$_{{Earth}_{s/n}}$}) and {\itshape L} = {\itshape R$_{E/SC}$} at the spacecraft ({\itshape L$_{s/c}$}). Then the integral $I_{1}$ can be written as

   \begin{displaymath}
    I_{1} = B \ R_\odot^{\epsilon}  \int_{- R_{S/E} \ cos \ \alpha} ^{R_{E/SC} - R_{S/E} \ cos \ \alpha} \ \frac{dx}{(x^2 + a^2)^{\epsilon/2}} 
 \end{displaymath}
 
 \begin{displaymath}
   \ \ \ = \frac {B \ R_\odot^{\epsilon}}{a^{\epsilon}} \int_{- R_{S/E} \ cos \ \alpha} ^{R_{E/SC} - R_{S/E} \ cos \ \alpha} \ \frac{dx}{(1 + \frac{x^2}{a^2})^{\epsilon/2}} \ \ .
 \end{displaymath}

      Now let
   \begin{displaymath}
   \frac{x}{a} = tan \ \theta \ \ ,
    \end{displaymath}
   and
   \begin{displaymath}
   dx = a\ sec^2 \ \theta \ d\theta \ \ .
   \end{displaymath}
   
   Therefore,
   \begin{equation}
   I_{1} = \frac {B \ R_\odot^{\epsilon}}{a^{\epsilon}} \int_{arctan\frac{(- R_{S/E} \ cos \ \alpha)}{a}} ^{arctan\frac{(R_{E/SC} - R_{S/E} \ cos \ \alpha)}{a}} \ \frac{a \ sec^2 \ \theta}{(tan^2\ \theta +1)^{\epsilon/2}} \ d\theta \ \ .
   \end{equation}

  From the geometry of Figure 3, the lower limit of Equation A.3 can be written as

  \begin{displaymath}
  arctan\bigg(\frac{- R_{S/E} \ cos \ \alpha}{a}\bigg) = arctan\bigg(\frac{- R_{S/E} \ cos \ \alpha}{R_{S/E} \ sin \ \alpha}\bigg) 
   \end{displaymath}
   \begin{displaymath}
  \ \ \ \ \ \ \ \ \ \ \ \ \ \ \ \ \ \ \ \ \ \  \ \ \ \ \ \ \ \ \ \ \ \ \ = arctan\bigg(-cot \ \alpha \bigg) \ \ ,
  \end{displaymath}

  with
  \begin{displaymath}
  cot \ \alpha = tan \ \bigg({\frac{\pi}{2} - \alpha}\bigg) \ \ .
  \end{displaymath}
  Hence, 
  \begin{displaymath}
arctan\bigg(\frac{- R_{S/E} \ cos \ \alpha}{a}\bigg) = \alpha - \frac{\pi}{2} \ \ .
    \end{displaymath}
    
    Similarly, the upper limit of Equation A.3 can be written as
    \begin{displaymath}
  arctan\bigg(\frac{R_{E/SC} - R_{S/E} \ cos \ \alpha}{a}\bigg) = arctan\bigg(\frac{R_{E/SC} - R_{S/E} \ cos \ \alpha}{R_{S/E} \ sin \ \alpha}\bigg) \ \ ,
   \end{displaymath}
   with
  \begin{displaymath}
  R_{E/SC} - R_{S/E} \ cos \ \alpha = R_{S/E} \ sin \ \alpha \ \bigg[\ tan \ \bigg( \ \beta - \bigg\{ \ \frac{\pi}{2} - \alpha \bigg\} \ \bigg) \ \bigg] \ \ .
   \end{displaymath} 
   Hence,
   \begin{displaymath}
  arctan\bigg(\frac{R_{E/SC} - R_{S/E} \ cos \ \alpha}{a}\bigg) = \beta + \alpha - \frac{\pi}{2} \ \ .
   \end{displaymath}

   Now Equation A.3 is given by
   \begin{displaymath}
   I_{1} = \frac {B \ R_\odot^{\epsilon}}{a^{\epsilon}} \int_{\alpha - \frac{\pi}{2}} ^{ \beta +  \alpha - \frac{\pi}{2}} \ \frac{a \ sec^2 \ \theta}{(tan^2\ \theta +1)^{\epsilon/2}} \ d\theta \ \ .
   \end{displaymath}
  
   with
   
   \begin{displaymath}
   sec^2 \ \theta = tan^2 \ \theta +1 \ \ ,
   \end{displaymath}
   and
   \begin{displaymath}
   cos \ \theta = \frac{1}{sec \ \theta} \ \ .
   \end{displaymath}
   
   Therefore, the integral $I_{1}$ can be written as
   \begin{equation}
 I_{1} = \frac {B \ R_\odot^{\epsilon}}{a^{\epsilon-1}} \int_{\alpha - \frac{\pi}{2}} ^{\beta + \alpha - \frac{\pi}{2}} (cos \ \theta)^{\epsilon - 2} \ d\theta \ \ .
   \end{equation}
   
   The maximum contribution of the integral occurs at $\theta$ = $0$. To solve this integral, Taylor series expansion was used and for $\theta$ near zero, it can be given as
   \begin{displaymath}
   f(\theta) = f(0) \ \ + \ \ \theta \ \bigg(\frac{df}{d\theta}\bigg) + \ \  \frac{\theta^2}{2!} \ \bigg(\frac{d^2f}{d\theta^2}\bigg) + \ \ \frac{\theta^3}{3!} \ \bigg(\frac{d^3f}{d\theta^3}\bigg)
   \end{displaymath}
   \begin{equation}
  \ \ \ \ \ \ \ \ \ + \ \frac{\theta^4}{4!} \ \bigg(\frac{d^4f}{d\theta^4}\bigg)  .......... + \ \theta \ (\theta^n) \ \ ,
   \end{equation}
   
   with
  \begin{displaymath}
   f(\theta) = (cos \ \theta)^{\epsilon - 2} \ \ .
    \end{displaymath}

   Then   
   \begin{displaymath}
  \ \ \ \ f(0) = 1 \ \ 
    \end{displaymath}
    
    \begin{displaymath}
    \bigg(\frac{df}{d\theta}\bigg) = -(\epsilon - 2) \ (sin \ \theta) \ (cos \ \theta)^{(\epsilon - 3)} \ \ ,
    \end{displaymath}
   \begin{displaymath}
   \bigg(\frac{df}{d\theta}\bigg)_{\theta=0} = 0 \ \ ,
   \end{displaymath}

    \begin{displaymath}
    \bigg(\frac{d^2f}{d\theta^2}\bigg) = (\epsilon - 2) \ (\epsilon - 3) \ (sin^2 \ \theta) \ (cos \ \theta)^{(\epsilon - 4)} 
    \end{displaymath}
   \begin{displaymath}
    \ \ \ \ \ \ \ \ \ \ \ \ \ - (\epsilon - 2)\ (cos \ \theta)^{(\epsilon - 2)} \ \ ,
    \end{displaymath}
   \begin{displaymath}
   \bigg(\frac{d^2f}{d\theta^2}\bigg)_{\theta=0} = -(\epsilon - 2) \ \ ,
   \end{displaymath}

   \begin{displaymath}
   \bigg(\frac{d^3f}{d\theta^3}\bigg) = -(\epsilon - 2) \ (\epsilon - 3) \ (\epsilon - 4) \ (sin^3 \ \theta) \ (cos \ \theta)^{(\epsilon - 5)}  
   \end{displaymath}
   \begin{displaymath}
   \ \ \ \ \ \ \ \ \ \ \ \ \ \ + (\epsilon - 2) \ (3\epsilon - 8) \ (sin \ \theta) \ (cos \ \theta)^{(\epsilon - 3)} \ \ ,
   \end{displaymath}
   \begin{displaymath}
   \bigg(\frac{d^3f}{d\theta^3}\bigg)_{\theta=0} = 0 \ \ ,
   \end{displaymath}

   \begin{displaymath}
   \bigg(\frac{d^4f}{d\theta^4}\bigg) = (\epsilon - 2) \ (\epsilon - 3) \ (\epsilon - 4) \ (\epsilon - 5)\ (sin^4 \ \theta) \ (cos \ \theta)^{(\epsilon - 6)} 
   \end{displaymath}
   \begin{displaymath}
   \ \ \ \ \ \ \ \ \ \ \ \ \ \ - (\epsilon - 2) \ (\epsilon - 3) \ (6\epsilon - 20) \ (sin^2 \ \theta) \ (cos \ \theta)^{(\epsilon - 4)} 
   \end{displaymath}
   \begin{displaymath}
   \ \ \ \ \ \ \ \ \ \ \ \ \ \ + (\epsilon - 2) \ (3\epsilon - 8) \ (cos \ \theta)^{(\epsilon - 2)} \ \ ,
   \end{displaymath}
   \begin{displaymath}
   \bigg(\frac{d^4f}{d\theta^4}\bigg)_{\theta=0} = (\epsilon - 2) \ (3\epsilon - 8) \ \ .
   \end{displaymath}

   Now, Equation A.5 can be written as
   
   \begin{displaymath}
   (cos \ \theta)^{(\epsilon - 2)} = 1 - {(\epsilon - 2)} \ \frac{\theta^2}{2!} + (\epsilon - 2) \ (3\epsilon - 8)\ \frac{\theta^4}{4!} + .... \ \theta \ (\theta^n)  
   \end{displaymath}
   \begin{displaymath}
   \ \ \ \ \ \ \ \ \ \ \ \ \ \ \ \ \ \ \ = 1 - \frac{\epsilon - 2}{2} \ \theta^2 + \frac{3\epsilon^2 -14\epsilon + 16}{24} \ \theta^4 + .... \ \theta \ (\theta^n) \ \ .
   \end{displaymath}
   
  By neglecting the higher order terms, the integral (Equation A.4) can be written as
  
   \begin{displaymath}
   I_{1} = \frac {B \ R_\odot^{\epsilon}}{a^{\epsilon-1}} \int_{\alpha - \frac{\pi}{2}} ^{\beta + \alpha - \frac{\pi}{2}} \bigg(1 - \frac{\epsilon - 2}{2} \    \theta^2 + \frac{3\epsilon^2 -14\epsilon + 16}{24} \ \theta^4 \bigg) \ d\theta 
   \end{displaymath} 
   \begin{displaymath}
   \ \ \ \ \ \ = \frac {B \ R_\odot^{\epsilon}}{a^{\epsilon-1}} \ \bigg[ \theta - \frac{\epsilon - 2}{6} \ \theta^3 + \frac{3\epsilon^2 -14\epsilon + 16}{120} \ \theta^5 \bigg]_{\alpha - \frac{\pi}{2}}^{\beta + \alpha - \frac{\pi}{2}} 
   \end{displaymath}   
   
   \begin{displaymath}
   \ \ \ \ \ \ = \frac {B \ R_\odot^{\epsilon}}{a^{\epsilon-1}}  \ \bigg[ \beta - \frac{\epsilon - 2}{6} \ \bigg( (\beta + \alpha - \pi/2)^3 - (\alpha - \pi/2)^3 \bigg)
   \end{displaymath} 
   \begin{displaymath}
   \ \ \ \ \ \ + \ \frac{3\epsilon^2 -14\epsilon + 16}{120} \ \bigg( (\beta + \alpha - \pi/2)^5 - (\alpha - \pi/2)^5 \bigg) \bigg] \ \ .
   \end{displaymath}
   
   By substituting the value of $a$, we can write the integral $I_{1}$ as
    \begin{displaymath}
    I_{1} = \frac {B \ R_\odot^{\epsilon}}{(R_{S/E} \ sin \alpha)^{\epsilon-1}}  \ \bigg[ \beta - \frac{\epsilon - 2}{6} \ \bigg( (\beta + \alpha - \pi/2)^3 - (\alpha - \pi/2)^3 \bigg)
   \end{displaymath} 
   \begin{equation}
   \ \ \ \ \ \ + \ \frac{3\epsilon^2 -14\epsilon + 16}{120} \ \bigg( (\beta + \alpha - \pi/2)^5 - (\alpha - \pi/2)^5 \bigg) \bigg] \ \ .
   \end{equation}

   Now, Equation A.2 can be written as
   $I_{2}$= $I_{2a}$ + $I_{2b}$
   where,
   \begin{equation}
   I_{2a} = \int_{L_{{Earth}_{s/n}}} ^{L_{s/c}} A\bigg(\frac{l}{R_\odot}\bigg)^{-4} \ dL \ \ ,
   \end {equation}
   and
   \begin{equation}
   I_{2b} = \int_{L_{{Earth}_{s/n}}} ^{L_{s/c}} B\bigg(\frac{l}{R_\odot}\bigg)^{-2} \ dL \ \ .
   \end {equation}
   
   Using a similar approach to the previous integral, one can write
   \begin{displaymath}
    I_{2a} = \frac {A \ R_\odot^{4}}{a^{3}} \int_{\alpha - \frac{\pi}{2}} ^{\beta + \alpha - \frac{\pi}{2}} cos^{2} \ \theta \ d\theta 
    \end{displaymath}
     \begin{displaymath}
    \ \ \ \ \ \ = \frac {A \ R_\odot^{4}}{a^{3}} \bigg[ \frac{1}{4}(2\ \theta + sin \ 2\theta)\bigg]_{\alpha - \frac{\pi}{2}}^{\beta + \alpha - \frac{\pi}{2}} 
   \end{displaymath}
   \begin{displaymath}
    \ \ \ \ \ \ = \frac {A \ R_\odot^{4}}{4 \ a^{3}} \bigg[ (2\ \beta - sin \ 2(\alpha + \beta) + sin \ 2\alpha)\bigg]
     \end{displaymath}
      \begin{displaymath}
    \ \ \ \ \ \ = \frac {A \ R_\odot^{4}}{4 \ R_{S/E}^{3} \ sin^{3} \alpha} \bigg[ (2\ \beta - sin \ 2(\alpha + \beta) + sin \ 2\alpha)\bigg] \ \ .
     \end{displaymath}
   
   Similarly, Equation A.8 can be written as
   \begin{displaymath}
    I_{2b} = \frac {B \ R_\odot^{2}}{a} \int_{\alpha - \frac{\pi}{2}} ^{\beta + \alpha - \frac{\pi}{2}} \ d\theta 
   \end{displaymath}
   \begin{displaymath}
    \ \ \ \ \ \ = \frac {B \ R_\odot^{2}}{a} \bigg[ \theta\bigg]_{\alpha - \frac{\pi}{2}}^{\beta + \alpha - \frac{\pi}{2}}
   \end{displaymath}
   \begin{displaymath}
    \ \ \ \ \ \ = \frac {B \ R_\odot^{2}}{R_{S/E} \ sin \ \alpha} \ \beta \ \ .
    \end{displaymath}
   
   Now by using $I_{2a}$ and $I_{2b}$, expression $I_{2}$ can be written as
       \begin{displaymath}
    I_{2} = \frac {A \ R_\odot^{4}}{4 \ R_{S/E}^{3} \ sin^{3} \alpha} \bigg[ (2\ \beta - sin \ 2(\alpha + \beta) + sin \ 2\alpha)\bigg] + \frac {B \ R_\odot^{2}}{R_{S/E} \ sin \ \alpha} \ \beta \ \ .
     \end{displaymath}
     \begin{equation}
     \end{equation}
  \end{appendix}    
\end{document}